\documentclass{aa}  

\usepackage{graphicx}
\usepackage{txfonts}
\usepackage{colortbl}
\usepackage{color,soul}
\usepackage[dvipsnames]{xcolor}

\newcommand{\um}{\hbox{\,$\mu$m}}
\newcommand{\av}{\hbox{A$_\mathrm{V}$}}
\newcommand{\msun}{\hbox{M$_\odot$}}
\newcommand{\kms}{\hbox{km\,s$^{-1}$}}
\definecolor{cinza}{rgb}{0.85,0.85,0.85} 
\usepackage{bm}
\usepackage{multirow}

\usepackage{hyperref}

\makeatletter
\renewcommand*\aa@pageof{, page \thepage{} of \pageref*{LastPage}}
\makeatother

\usepackage{dashbox,framed,color,ocg-p}
\begin{document}

   \title{A wide survey for circumstellar disks in the Lupus complex}
   \titlerunning{Lupus as part of the Sco-Cen complex}

   \author{P.~S. Teixeira
          \inst{1}
          \and
          A. Scholz
          \inst{1}
          \and
          J. Alves
          \inst{2}
          }

   \institute{Scottish Universities Physics Alliance (SUPA), School of Physics and Astronomy, University of St. Andrews, North Haugh, Fife, KY16 9SS, St. Andrews, UK\\
              \email{psdvt@st-andrews.ac.uk}
     \and
     University of Vienna,Department of Astrophysics, T\"urkenschanzstrasse 17, A-1180 Vienna, Austria \\
             }

   \date{}

 
  \abstract
   {
   Previous star formation studies have, out of necessity, often defined a population of young stars confined to the proximity of a molecular cloud. Gaia allows us to examine a wider, three-dimensional structure of nearby star forming regions, leading to a new understanding of their history.
   We present a wide-area survey covering 494\,deg$^2$ of the Lupus complex, a prototypical low-mass star forming region. Our survey includes all known molecular clouds in this region as well as parts of the Upper Scorpius (US) and Upper Centaurus Lupus (UCL) groups of the Sco-Cen complex.
   We combine Gaia DR2 proper motions and parallaxes as well as ALLWISE mid-infrared photometry to select young stellar objects (YSOs) with disks. 
   The YSO ages are inferred from Gaia color-magnitude diagrams, and their evolutionary stages from the slope of the spectral energy distributions.
   We find 98 new disk-bearing sources. Our new sample includes objects with ages ranging from 1 to 15\,Myr and masses ranging from 0.05 to 0.5$\,M_{\odot}$, and consists of 56 sources with thick disks and 42 sources with anemic disks.
   While the youngest members are concentrated in the clouds and at distances of 160\,pc, there is a distributed population of slightly older stars that overlap in proper motion, spatial distribution, distance, and age with the Lupus and UCL groups. 
   The spatial and kinematic properties of the new disk-bearing YSOs indicate that Lupus and UCL are not distinct groups.
  Our new sample comprises some of the nearest disks to Earth at these ages, and thus provides an important target for follow-up studies of disks and accretion in very low mass stars, for example with ALMA and ESO-VLT X-Shooter. }

   \maketitle
%

\section{Introduction}
\label{sec:intro}

\noindent The highly successful \emph{Spitzer} mid-infrared observational studies of young stellar clusters have been mostly spatially restricted to regions of dense molecular clouds because it was too expensive and/or not feasible to explore larger areas on the sky \citep[e.g.,][]{evans09}. As a result, the extended population of young stellar clusters were not fully sampled or surveyed. Data from all-sky surveys are needed to identify and characterize these extended populations. Young stellar objects (YSOs) that populate the extended haloes of star formation regions can be identified with the Wide-field Infrared Survey Explorer cryogenic survey \citep[WISE;][]{wright10}, particularly for nearby star forming regions where WISE is sensitive enough to sample the majority of the initial mass function (IMF) mass range. Furthermore, before the advent of the revolutionary Gaia astrometric mission \citep{gaia16}, even most of the nearby star formation regions had insufficient proper motion information available to establish kinematic membership down to low stellar masses.
The combination of these two multiwavelength all-sky surveys, WISE and Gaia, are allowing us to now complete cluster membership, which is essential for obtaining more accurate values of, for example, the IMF, disk fractions, star formation rates, star formation history, and cluster expansion rate. This paper analyzes the extended population of the Lupus complex as a demonstration of the importance of identifying off-cloud stellar members of young stellar clusters. Lupus  has become a benchmark for studies of disks and accretion \citep[e.g.,][]{merin08,alcala17} and also for studies of brown dwarf formation \citep[e.g.,][]{muzic14}. In light of new data availability, it is now necessary to revisit this region with a broader perspective to update its membership list and star formation history.\\

\begin{figure}[!h]
\centering
\includegraphics[width=0.5\textwidth]{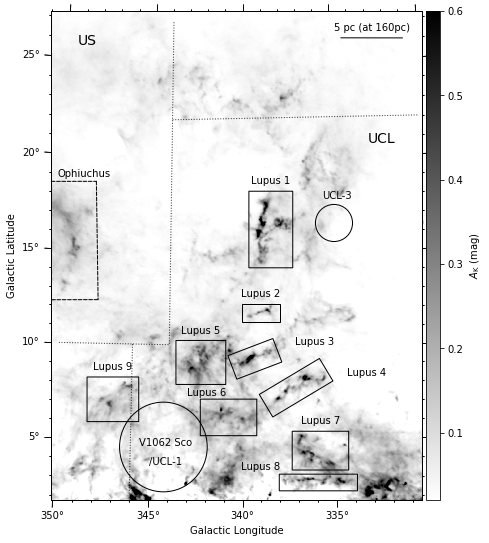}
\caption{Planck extinction map of the Lupus complex \citep{planck11}, showing the area surveyed in this paper. The eight historically identified Lupus clouds are marked by rectangles. A scale bar of 5\,pc (at a distance of 160\,pc) is shown for reference. Part of the Ophiuchus cloud is located within the surveyed area, as identified by the dashed rectangle. The dotted lines mark the borders of Upper Scorpius and Upper Centaurus-Lupus subgroups according to \citet{dezeeuw99}. The circles mark the locations of the compact groups V1062\,Sco/UCL-1 and UCL-3  \citep{roeser18, damiani19}.}
\label{fig:clouds}
\end{figure}

\noindent  The Lupus complex is conventionally described as being composed of nine clouds \citep{hara99}. The distance to the individual clouds vary between 200\,pc and 140\,pc \citep{zucker19}. The complex spans an area of  roughly 22\degr$\times$15\degr, which corresponds to a physical size of approximately 62$\times$42\,pc, and sits just above the Galactic plane. 
Figure \ref{fig:clouds} shows an extinction map of the complex in which the nine clouds are indicated by rectangles. 
The clouds are different in terms of their star formation activity; the most productive cloud is Lupus\,3, which has an optically revealed stellar cluster of young accreting stars \citep{schwartz77, nakajima03,comeron03}. \emph{Spitzer} observations have confirmed the relatively high fraction of YSOs in Lupus\,3 compared to the other clouds \citep{cieza07b,merin08}. Even though Lupus\,3 has a more evolved YSO population, it also has dense cores and protostars still forming within those cores \citep{teixeira05,tachihara07}. The other Lupus clouds that also have dense protostellar cores are Lupus\,1 and Lupus\,4  \citep{benedettini11,rygl13}. The remaining clouds in the Lupus complex, namely, Lupus\,2, Lupus\,5, Lupus\,6, Lupus\,7, Lupus\,8, and Lupus\,9, do not appear to have ongoing star formation; however, it should be noted that these latter clouds have not been observed in the same level of detail as Lupus\,1, Lupus\,3, and Lupus\,4. Furthermore, currently little is known about the off-cloud Lupus population.\\

\noindent The area surveyed in this paper (Fig. \ref{fig:clouds}) also includes part of the Upper Scorpius (US; age $\approx$ 4-6\,Myr) and the Upper Centaurus-Lupus (UCL; age $\approx$ 15-22\,Myr) subgroups, and the co-moving subgroup V1062 Scorpii (V1062Sco; age $<$ 25\,Myr) in the Scorpius-Centaurus OB association \citep[][]{dezeeuw99,mamajek02, wright18, roeser18}. These populations, particularly those of Lupus and UCL, therefore overlap along the line-of-sight, which is the main limitation when searching for off-cloud Lupus YSO members. Indeed, disentangling different populations along the line-of-sight is an issue that is increasingly important to solve as survey areas get larger. In this particular case, because the distances spanned by Lupus, UCL, US, and V1062\,Sco also overlap, these populations may be physically connected \citep[as hinted by ][]{comeron08}. \\

\noindent A recent analysis of the Sco OB2 association carried out by \citet{damiani19} using the second data release of Gaia has identified several populations within the region shown in Figure\,\ref{fig:clouds}, namely, the compact groups UCL-1 (corresponding to V1062\,Sco), UCL-3, and Lup-III, and the diffuse groups D1 and D2 (further subdivided into sub-populations D2a, D2b, and USC-D2 in Upper Sco). We revisit the region encompassing the Lupus molecular cloud complex to identify the dispersed disk population associated with the Lupus molecular clouds.
\\

\noindent We make use of the two aforementioned complementary all-sky photometric surveys to identify disk-bearing YSOs: ALLWISE \citep{cutri13} in the mid-infrared to identify sources with excess emission characteristic of circumstellar disks, and Gaia \citep{gaia16} to identify sources that share the same proper motion and distance with the aim to attribute kinematic membership of new disked sources  to Lupus, UCL, US, or V1062\,Sco. The paper is organized as follows: \S\,\ref{sec:data} describes these two datasets and the flagging performed to remove possible contaminants; \S\,\ref{sec:selection} shows the selection criteria used for identifying YSOs; \S\,\ref{sec:charaterization} explores their mid-infrared and astrometric properties; \S\,\ref{sec:discussion} explores the results, and finally we present the conclusions in \S\,\ref{sec:conclusion}.

\section{Datasets and filtering}
\label{sec:data}

The area analyzed in this paper covers the entire Lupus complex, extending from \hbox{330\degr $< l <$ 349\degr}, and \hbox{1.6\degr $< b <$  27.6\degr}, as shown in Figure \ref{fig:clouds}, including all the nine Lupus clouds and an extensive area around them. The purpose of using such a large area is to identify both on- and off-cloud sources in a systematic and unbiased manner to test the boundaries of individual star forming regions, to ultimately achieve a complete census of well characterized YSOs, and to constrain the star formation history for low-mass regions.

\subsection{Gaia}
\noindent Gaia \citep{gaia16} data release 2 \citep[DR2,][]{gaiadr218, lindegren18} data was obtained through the ESA Gaia TAP service in TOPCAT \citep{taylor05}. The Astronomical Data Query Language (ADQL) \citep{osuna08} script is shown in Appendix \S\,\ref{sec:gaiaquery}. We restricted the query to sources with parallaxes between 4\,mas and 12\,mas (covering the distance range to the Lupus clouds found by \citet{zucker19}), and with errors in their proper motions to less than 1\,mas. In this work we adopt the Gaia distance estimates calculated by \citet{bailer-jones18}.

\subsection{ALLWISE}
\noindent In addition to Gaia data, we use ALLWISE photometry \citep{cutri13} to identify sources with excess infrared emission. AllWISE makes use of data from the Wide-field Infrared Survey Explorer cryogenic survey \citep[WISE][]{wright10} and NEOWISE post-cryogenic survey \citep{mainzer11}; it consists of mid-infrared photometry in four broad bands, W1, W2, W3, and W4, centered at wavelengths 3.6\um, 4.6\um, 11.6\um, and 22.1\um, respectively. The ESA Gaia archive\footnote{\url{https://www.cosmos.esa.int/web/gaia/data-release-2}} provides cross-matched catalogs between Gaia DR2 and other surveys, including ALLWISE. The ADQL query we used runs an identity cross-match using the Gaia source IDs between the catalogs \texttt{gaiadr2.gaia\_source} and \texttt {gaiadr2.allwise\_best\_neighbour}. 
Since the Lupus complex extends to very low Galactic latitudes, cross-matching Gaia with ALLWISE can be problematic due to crowding near the Galactic plane (exacerbated by WISE's relatively large PSF) and lead to missed matches. \citet{wilson17} have explored this issue in detail and provide an alternative cross-matched catalog for the Galactic plane region, |b| $<$ 10\degr \citep{wilson18}. The aforementioned cross-match between \texttt{gaiadr2.gaia\_source} and \texttt {gaiadr2.allwise\_best\_neighbour} for b\,$<$\,10\degr\ yields 8,\,210 sources. We queried the TAPVizieR table \texttt{IV/35/wn18\_b10} \citep{wilson18} (see Appendix \S\,\ref{sec:gaiaquery} for the ADQL script) and did a cross-match with the  \texttt{gaiadr2.gaia\_source} table using the Gaia source IDs and obtained 9,\,289 matches. We thus use the latter catalog for b\,$<$\,10\degr\ since it yielded 1,\,079 more matches. Finally, we used the NASA/IPAC Infrared Science Archive to obtain the ALLWISE photometry and respective flags, cross-matching on the ALLWISE source IDs. This combined catalog also contains near infrared photometry from 2MASS \citep{skrutskie06}.

\noindent Due to the large area of the Lupus complex,  and its proximity to the Galactic plane, it is necessary to filter out possible contaminants in our catalog. In order to achieve this we set four conditions on the ALLWISE data\footnote{see \url{http://wise2.ipac.caltech.edu/docs/release/allwise/expsup/sec2\_1a.html\#cc\_flags} for a complete description of the column descriptions in the ALLWISE source catalog.}. We first set \hbox{\emph{cc\_flags = 0000}}; the contamination and confusion flag is composed of a four-character string, one character per band [W1/W2/W3/W4], that indicates that the photometry and/or position measurements of a source may be contaminated or biased due to proximity to an image artifact. Values of zero indicate that the source is unaffected by known artifacts. We then set \hbox{\emph{ext\_flg = 0}}; the extended source flag is an integer flag that we use to exclude extended sources (e.g., galaxies and cirrus mis-identifications). A value of zero indicates that the source shape is consistent with a point-source and the source is not associated with or superimposed on a 2MASS Extended Source Catalog source. Some YSOs that are embedded and have reflection nebulae may be filtered out using this criterium.
Thirdly, we set \hbox{\emph{w1snr, w2snr, w3snr $>$ 5}}; a signal-to-noise ratio detection greater than 5 for the bands W1, W2, and W3.
Lastly, we set \hbox{\emph{w1flg, w2flg, w3flg $\leq$ 1}}; the aperture measurement quality flag indicates if one or more image pixels in the measurement aperture for this band is confused with nearby objects, is contaminated by saturated or otherwise unusable pixels, or is an upper limit. A value of zero indicates no contamination, while a value of 1 indicates that another source falls within the measurement aperture. We retain the latter sources in our YSO selection; the new YSOs with these values are appropriately identified in this paper. Removing them from the YSO sample does not change the main conclusions of the paper, so we opted to keep them in as sources worth further investigation in the future.

\noindent It should be noted that not all sources with disks in the area surveyed will be included in our YSO survey, either because the ALLWISE data is not sensitive enough to detect them or because the sources are removed by the aforementioned filtering conditions. We are therefore not able to provide a complete sample of disk-bearing sources, however, the sample we do provide is a robust list of bona fide disk-bearing sources, free from contaminants.

\subsection{Previously identified members}
\label{subsec:pkm}
\noindent We compiled a catalog of confirmed and high probability members of the Lupus clouds previously published in the literature, namely from \citet{merin08} and \citet{alcala17}. The list of members was cross-matched with the Gaia DR2 catalog and used as a reference point for the proper motion selection, as explained in the next section. 
The US members were obtained from \citet{luhman18}, and Upper Centaurus-Lupus members from \citet{dezeeuw99}; both member lists were cross-matched with the Gaia DR2 catalog. The V1062\,Sco members were obtained from \citet{roeser18}. Despite the fact that part of the Ophiuchus molecular cloud is included in our survey area (see Fig. \ref{fig:clouds}), we found no Gaia DR2 detected Ophiuchus members \citep[after cross-matching with the membership list from][]{canovas19}.\\

\noindent We found Gaia DR\,2 counterparts to 75\% of the aforementioned Lupus membership list. Of these matched sources, we found ALLWISE counterparts to 26\% of the diskless Lupus members, and to 65\% of the Lupus disk-bearing sources. After applying the ALLWISE filtering (see Section 2.2) to the sample of Lupus members with Gaia DR\,2 and ALLWISE counterparts, we recover 35\% of the known Lupus members with disks. The conservative ALLWISE filtering gives us greater confidence that we will select bona-fide disk-bearing sources.


\section{Selection of disk-bearing YSOs}
\label{sec:selection}

\paragraph{Spectral energy distribution slope\newline}

\noindent To identify YSOs with disks, we use the spectral energy distribution (SED) slope between the bands W1 and W3, following the widely used Lada classification scheme \citep{lada87, lada06a}. The W4 band was not used in this work to identify excess emission because the majority of the photometry in this band corresponds to either upper limits or is severely affected by source confusion. The SED slope is therefore defined as:

\begin{equation}
\alpha_{ALLWISE} = \frac{d log(\lambda F_{\lambda})}{d log(\lambda)},
\label{eq:slope}
\end{equation}

\noindent where $F_\lambda$ is the flux density for a given wavelength $\lambda$, and \hbox{$\lambda$ = [3.6, 4.6, 11.6]\,\um}.
 The sources were classified into five categories according to $\alpha_{ALLWISE}$, as described in Table \ref{tab:alpha_wise}, and this paper is centered on the analysis of the population of sources with Thick Disks (TDs) and Anemic Disks (ADs).

\begin{table}[!h]
\centering
\caption{YSO classification scheme.}
\label{tab:alpha_wise}
\centering
\begin{tabular}{lc}
\hline
Source classification & $\alpha_{ALLWISE}$\\
\hline
\hline
Photosphere & $\alpha_\mathrm{ALLWISE} < $ -2.3 \\
Anemic disk & -2.3 $ \leq\alpha_\mathrm{ALLWISE} \leq $ -1.8 \\
Thick disk & -1.8 $ < \alpha_\mathrm{ALLWISE} \leq $ -0.3 \\
Flat spectrum & -0.3$ < \alpha_\mathrm{ALLWISE} \leq $ 0.3 \\
Protostar & 0.3 $ < \alpha_\mathrm{ALLWISE}$ \\
\hline
\end{tabular}
\end{table}

\paragraph{Proper motion\newline}

\noindent Having identified disk-bearing YSOs within the region shown in Figure \ref{fig:clouds}, for distances between 110\,pc and 200\,pc, we then proceeded to identify those sources that have proper motions consistent with those of previously identified members of Lupus \citep[obtained from][]{merin08, alcala17}. 
The known Lupus members with Gaia DR\,2 counterparts have mean proper motion values of \hbox{$\mu_\alpha^* = -12\pm 3$\,mas} and \hbox{$\mu_\delta =  -23 \pm 2$\,mas}. 
We use a range of  $\pm$5$\sigma$ as proper motion cuts to select new Lupus YSOs, namely:  $-27 < \mu_\alpha^* < 3\ \mathrm{mas}$ and $-33 < \mu_\delta < -13\  \mathrm{mas}$.
\noindent After applying the proper motion selection across the entire region surveyed, we identified 98 new disk-bearing sources, of which 42 are sources with anemic disks and 56 are sources with thick disks.

\paragraph{Variability\newline}
\noindent Optical and mid-infrared variability may be an indicator of youth and is very commonly observed in YSOs \citep[e.g.,][]{alencar10,cody14}.
The ALLWISE catalog contains information about variability, derived from the several observing epochs of the NEOWISE campaign for bands W1 and W2. This information is given by the flag $"var\_ flg,"$ where we take values greater or equal than 6 to denote variability \citep{hoffman12}. 
Of the new disk-bearing YSOs, there is no variability information for 31 sources; for the remaining sources, 8 present variability in bands W1 and W2 (1 TD and 7 ADs).  The Gaia DR2 catalog also contains a variability flag, and we find 4 sources classified as variable  (1 TD and 3 ADs). However, for the majority of the new YSO sources there is no Gaia variability information available. We identified only one new YSO that presents variability in both Gaia and ALLWISE data. \\

\begin{figure}[!h]
    \centering
    \includegraphics[width=0.5\textwidth]{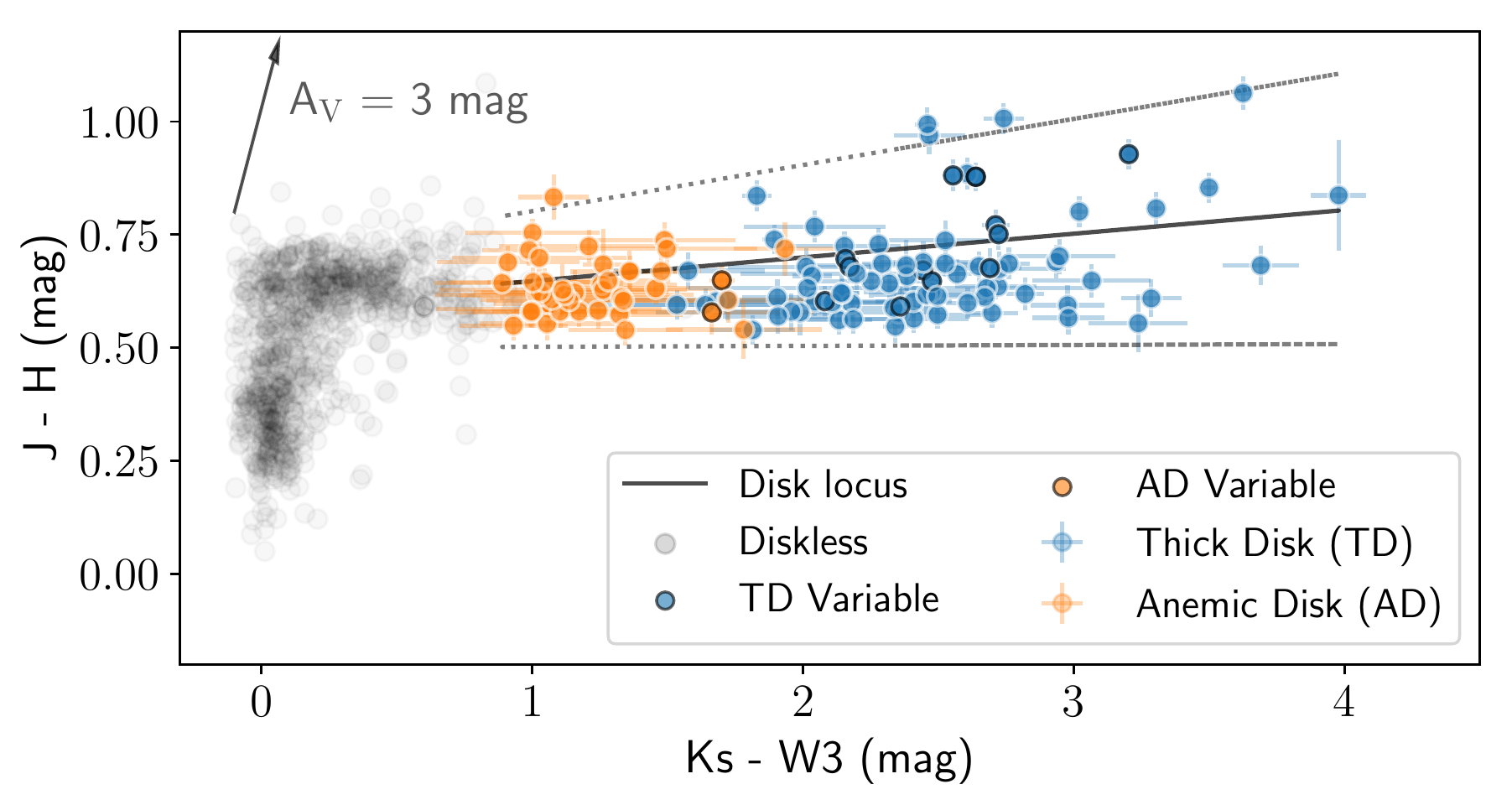}
    \caption{ 2MASS-WISE $K_s - W3\ vs.\ (J-H)$ color-color diagram of the new disked sources. The dotted lines denote the position of the disk locus $\pm 5\sigma$. The extinction vector was calculated using the extinction law from \citet{wang19}.}
    \label{fig:cc}
\end{figure}

\noindent  Figure \ref{fig:cc} shows a color-color diagram for the new disked sources. We calculated an empirical disk locus for the ${K_s - W3}\ versus\ {(J-H)}$ color-color space by performing a weighted least squares fit to the colors of the disked sources \citep[see][]{teixeira12}. The disk locus found is:
\begin{equation}
(J-H) = (0.05\pm 0.01) \cdot (K_s - [W3]) + (0.60 \pm 0.02).
\end{equation}

\section{Characterization of the new disk-bearing YSOs}
\label{sec:charaterization}

\subsection{Spatial distribution}

\begin{figure*}[!h]
\centering
\includegraphics[width=0.9\textwidth]{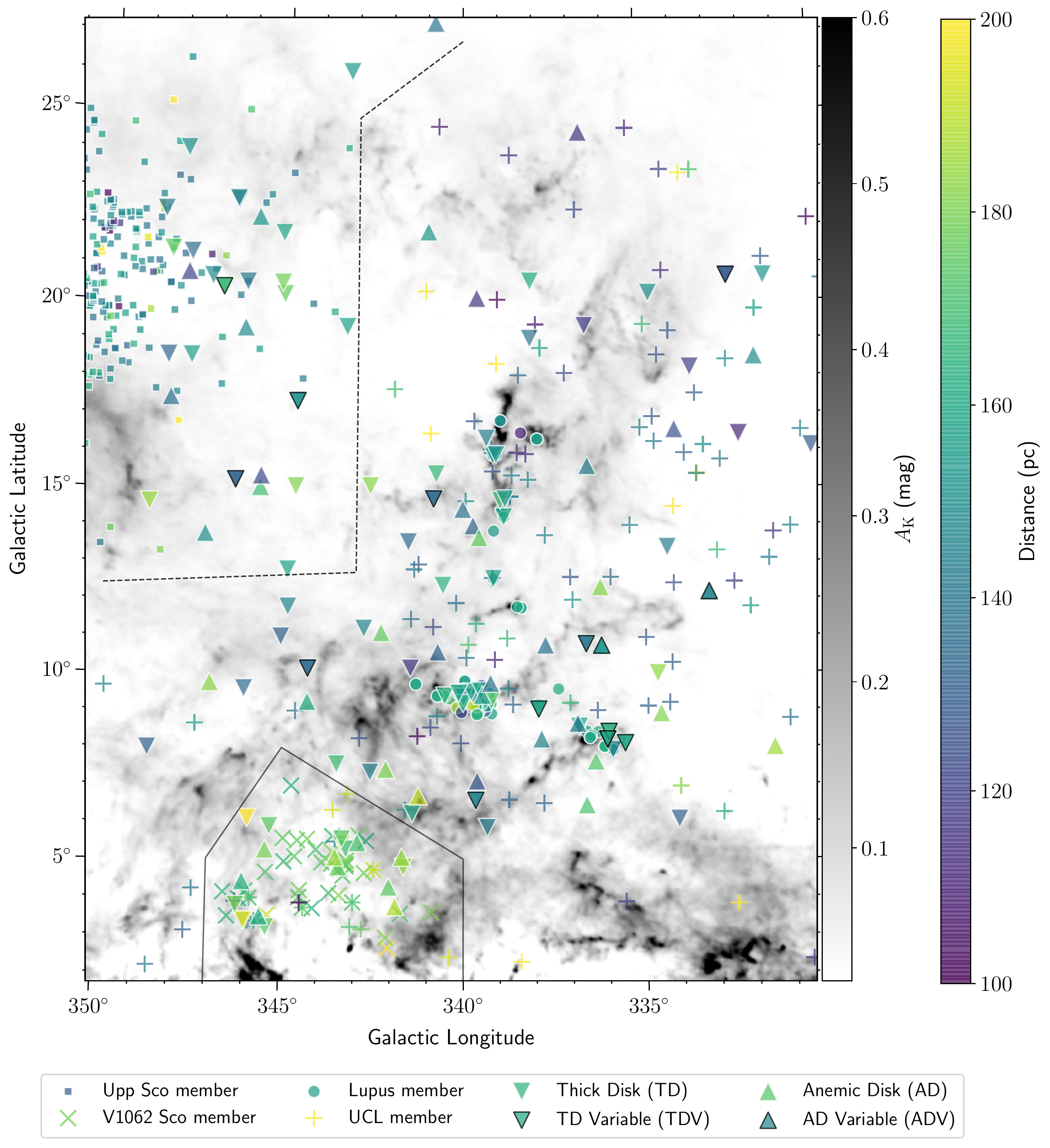}
\caption{Spatial distribution for the new disk-bearing YSOs (TD, TDV, AD, ADV), and previously known members of Upper Sco, V1062\,Sco, UCL, and Lupus, superimposed on the Planck \av\ map. The Upper Sco and V1062 Sco YSOs are separated from those of Lupus and/or UCL by the dashed and solid lines, respectively. A more detailed version of this plot is shown in Figure \ref{fig:s_dist_panel_appendix}, where the spatial distribution is shown for different distance slices.}
\label{fig:s_dist_panel}
\end{figure*}

Figure \ref{fig:s_dist_panel} shows how the disked population is spatially distributed throughout the region. The more distant YSOs tend to be located in between Lupus\,5, Lupus\,6, and Lupus\,9 and the Galactic Plane; these are likely part of the newly discovered sub-group V1062\,Scorpii/UCL-1 \citep{roeser18,damiani19}. Figure \ref{fig:s_dist_panel_appendix} shows the same distribution for slices of different distance bins.
For distances around 160$\pm$10\,pc, the sources are located near Lupus\,1, Lupus\,3, and Lupus\,4. The nearest YSOs occupy high Galactic Latitudes. This source distribution is consistent with the distances that \citet{zucker19} found for the Lupus region, specifically that the clouds closest to the Galactic Plane were more distant. Along with this spatial trend of the YSOs, there is a group of sources that populates the US (and Ophiuchus) region, mostly from 135\,pc - 160\,pc. Figures \ref{fig:s_dist_panel} and  \ref{fig:s_dist_panel_appendix} show that the previously known members of Lupus and UCL overlap spatially. The Lupus sources are all located near the molecular clouds, but so are some of the UCL sources.

\subsection{Gaia proper motion}
\begin{figure*}[!h]
\centering
\includegraphics[width=\textwidth]{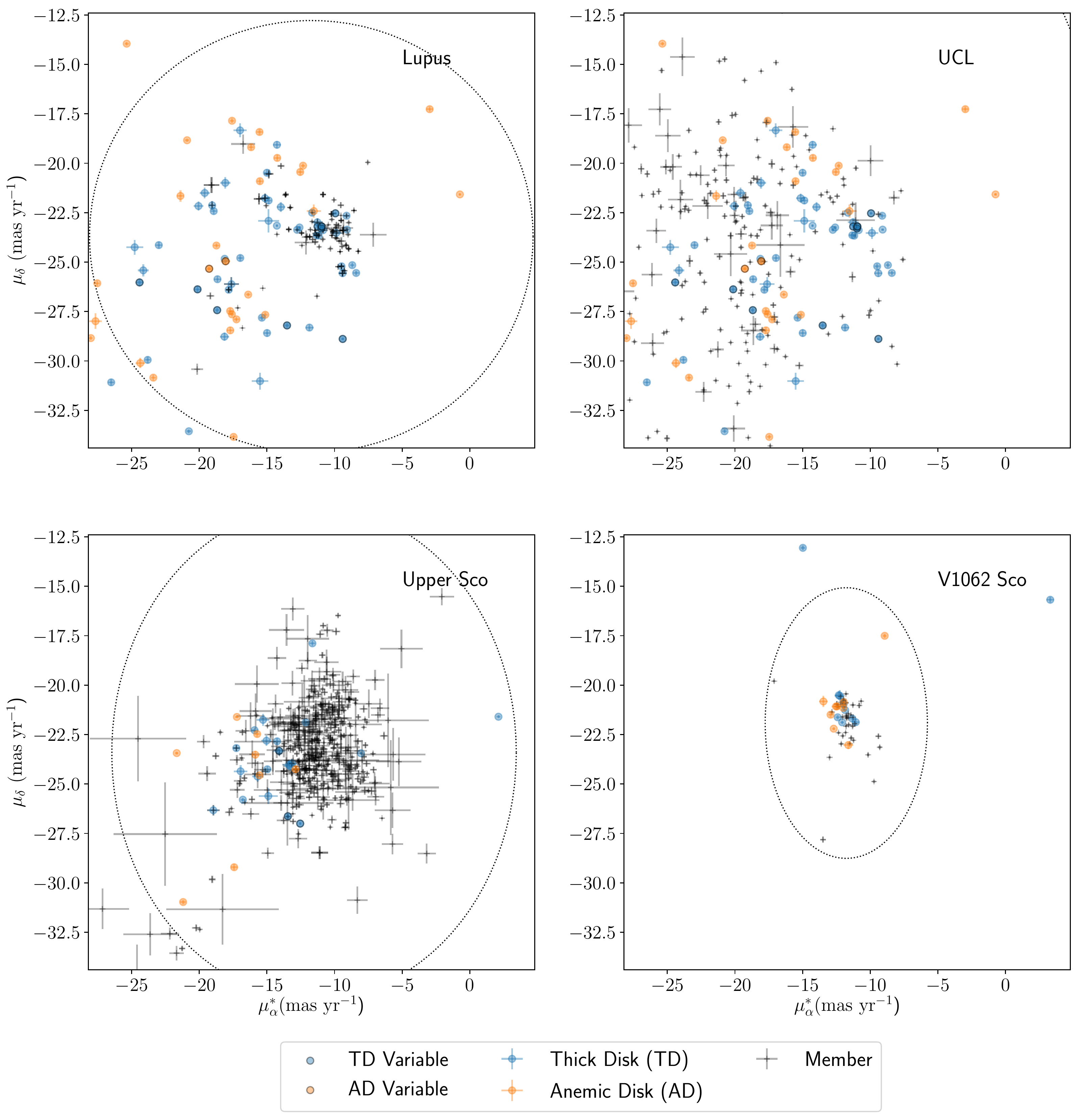}
\caption{Gaia proper motion diagrams for previously known YSO members and the new disk-bearing YSOs presented in this paper. Both upper panels show the same new  disk-bearing sources, whereas the bottom panels show the new disk-bearing sources restricted to the respective spatial locations of Upper Sco and V1062 Sco. All panels show an ellipse corresponding to 5\,$\sigma$ of the mean proper motion of the respective confirmed members (the UCL ellipse is larger than the panel). }
\label{fig:pm}
\end{figure*}

\begin{figure*}[!h]
\centering
\includegraphics[width=\textwidth]{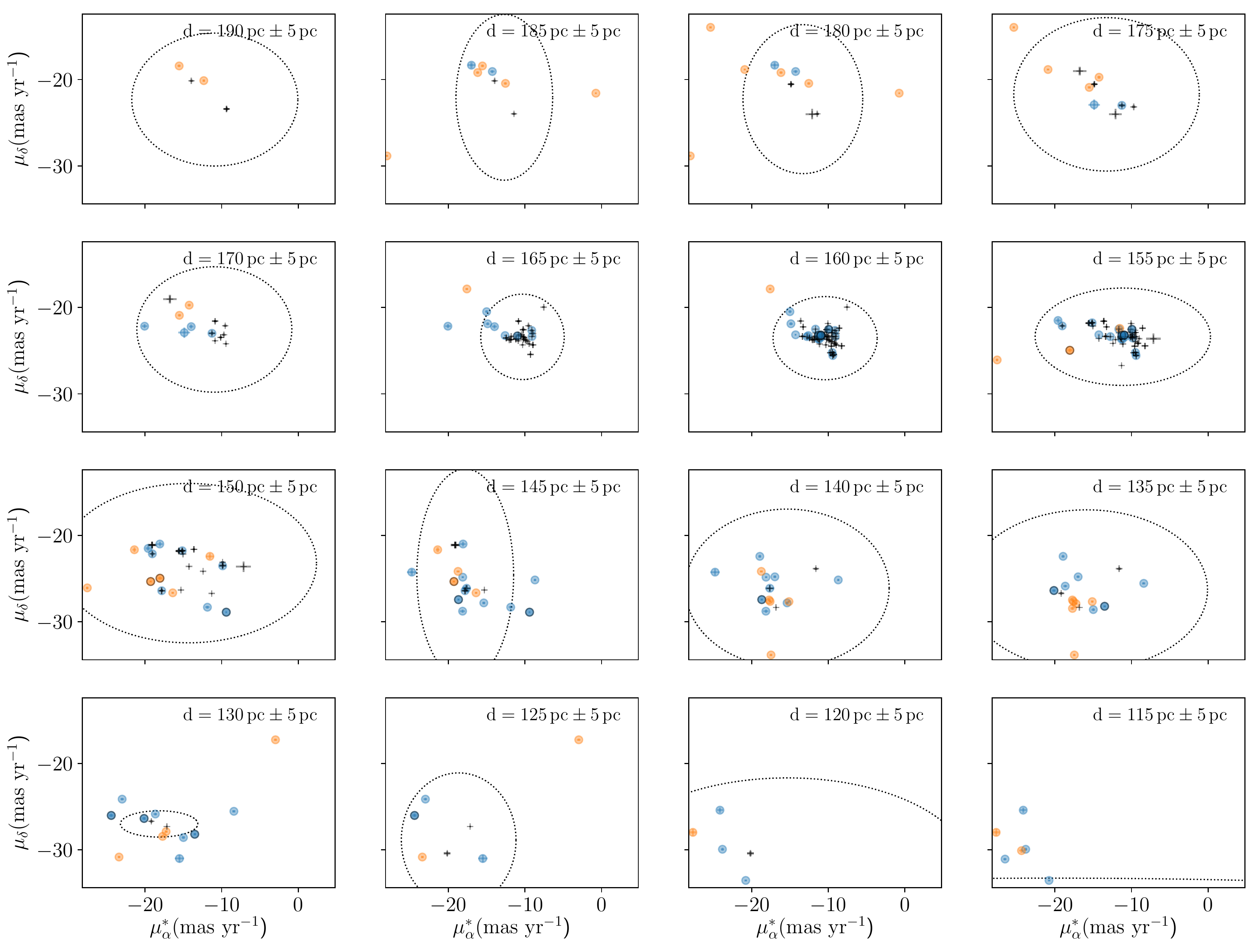}
\caption{Gaia proper motion diagrams for different distance slices, in steps of 5\,pc, Nyquist sampled, for the new disk-bearing YSOs and previously known Lupus members. The symbols are the same as those shown in Figure \ref{fig:pm}. Each panel shows the 5\,$\sigma$ proper motion ellipse for the known members at that distance.}
\label{fig:pm_panel}
\end{figure*}

\begin{figure*}[!h]
\centering
\includegraphics[width=\textwidth]{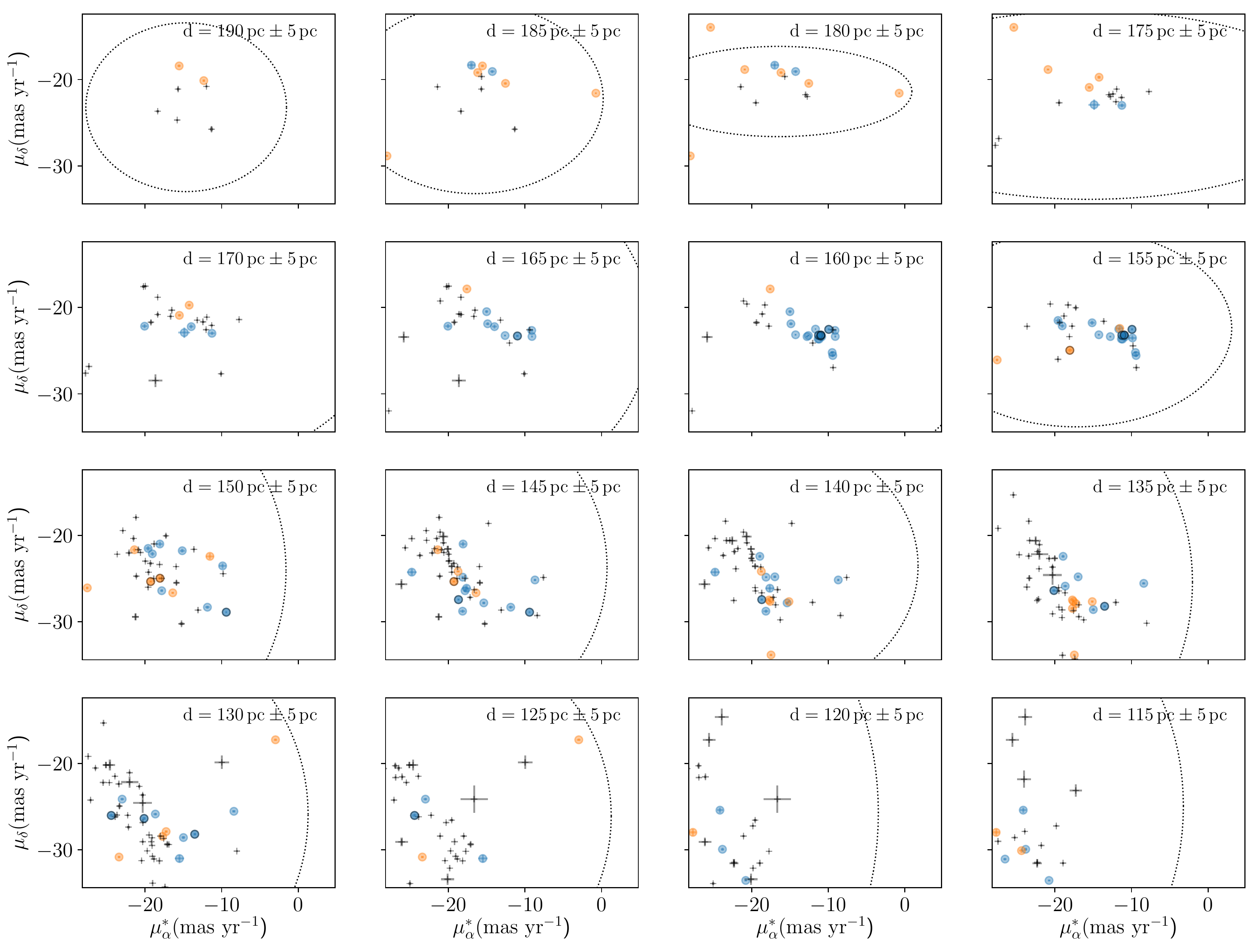}
\caption{Gaia proper motion diagrams for different distance slices, in steps of 5\,pc, Nyquist sampled, for the new disk-bearing YSOs and previously known UCL members. The color symbols are the same as those shown in Figure \ref{fig:pm}. Each panel shows the 5\,$\sigma$ proper motion ellipse for the known members at that distance.}
\label{fig:pm_panel_ucl}
\end{figure*}

\begin{figure*}[!h]
\centering
\includegraphics[width=\textwidth]{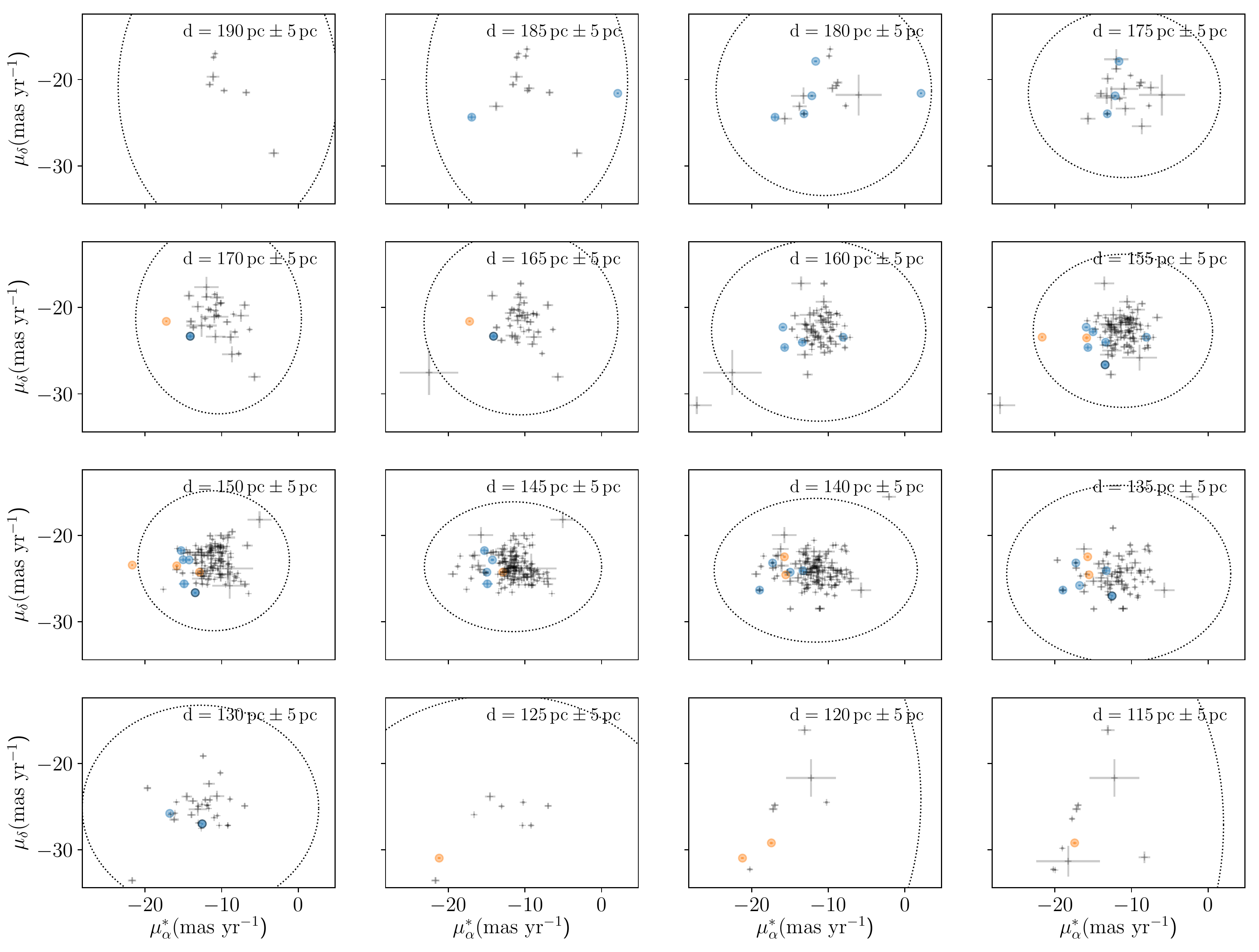}
\caption{Gaia proper motion diagrams for different distance slices, in steps of 5\,pc, Nyquist sampled, for the new and previously known Upper Scorpius YSOs. The symbols are the same as those shown in Figure \ref{fig:pm}. Each panel shows the 5\,$\sigma$ proper motion ellipse for the known members at that distance.}
\label{fig:pm_panel_uppsco}
\end{figure*}

\begin{figure*}[!h]
    \centering
    \includegraphics[width=\textwidth]{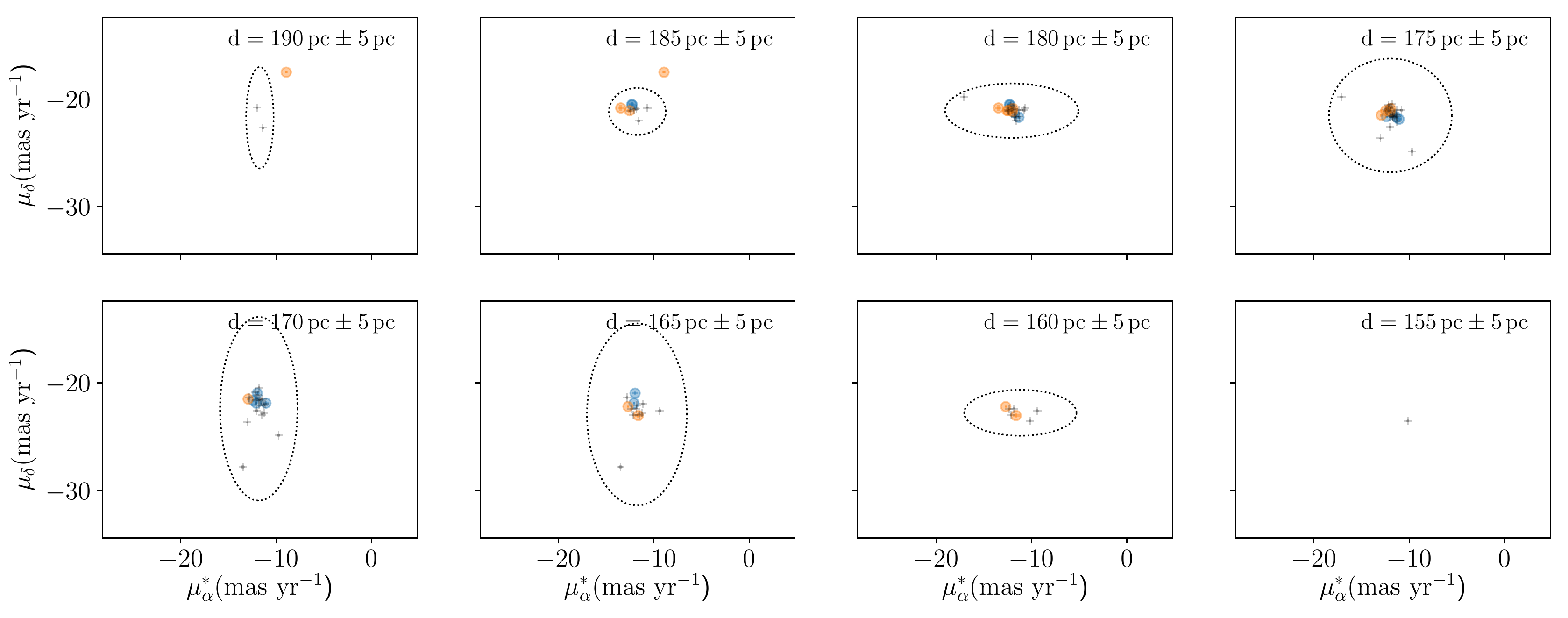}
    \caption{Gaia proper motion diagrams for different distance slices, in steps of 5\,pc, Nyquist sampled, for the new and previously known V1062\,Sco YSOs. The color symbols are the same as those shown in Figure \ref{fig:pm}. Each panel shows the 5\,$\sigma$ proper motion ellipse for the known members at that distance.}
    \label{fig:pm_panel_v1062sco}
\end{figure*}

\noindent As explained in Section \ref{sec:selection}, the new YSOs with disks have proper motions within 5-$\sigma$ of the proper motions of previously known members of Lupus. 
 The four panels of Figure \ref{fig:pm} show the proper motions of previously known YSO members (pkm) published in the literature for Lupus \citep{merin08, alcala17}, Upper Centaurus-Lupus \citep{dezeeuw99}, Upper Scorpius  \citep{luhman18}, V1062\,Sco \citep{roeser18}, and the new  YSOs for each stellar group: the upper left panel shows the proper motions for Lupus, the upper right panel shows the proper motions for UCL; both upper panels show the same new YSO sources. The lower panels shows the proper motions for US (left) and V1062\,Sco (right) and their respective new disk-bearing YSOs.
 The majority of the previously known members are located between 150\,pc and 170\,pc, yet there are members with distances spanning the entire range, from 115\,pc tp 190\,pc.  Since the sources span a relatively large distance range, the Lupus proper motion plot is reproduced for different distance intervals in Figure \ref{fig:pm_panel}; the distance step is larger than the median distance error for the sources. The sources are most concentrated at distances 155\,pc to 165\,pc, for larger and nearer distances there is more dispersion within the proper motion plots but the new disk-bearing YSOs still follow the Lupus members. 
 Figure \ref{fig:pm_panel_ucl}
shows a panel of proper motion plots for different distance slices for the known UCL members and the same new disk-bearing sources shown in Figure \ref{fig:pm_panel}. Figures \ref{fig:pm_panel} and \ref{fig:pm_panel_ucl} show that the new disk-bearing sources have proper motions consistent with both co-moving groups. We show for completeness the same panels of proper motions according to distance for US (Figure \ref{fig:pm_panel_uppsco}) and V1062\,Sco (Figure \ref{fig:pm_panel_v1062sco}).
 
\subsection{Gaia color-magnitude diagrams}

\begin{figure*}[!h]
\centering
\includegraphics[width=\textwidth]{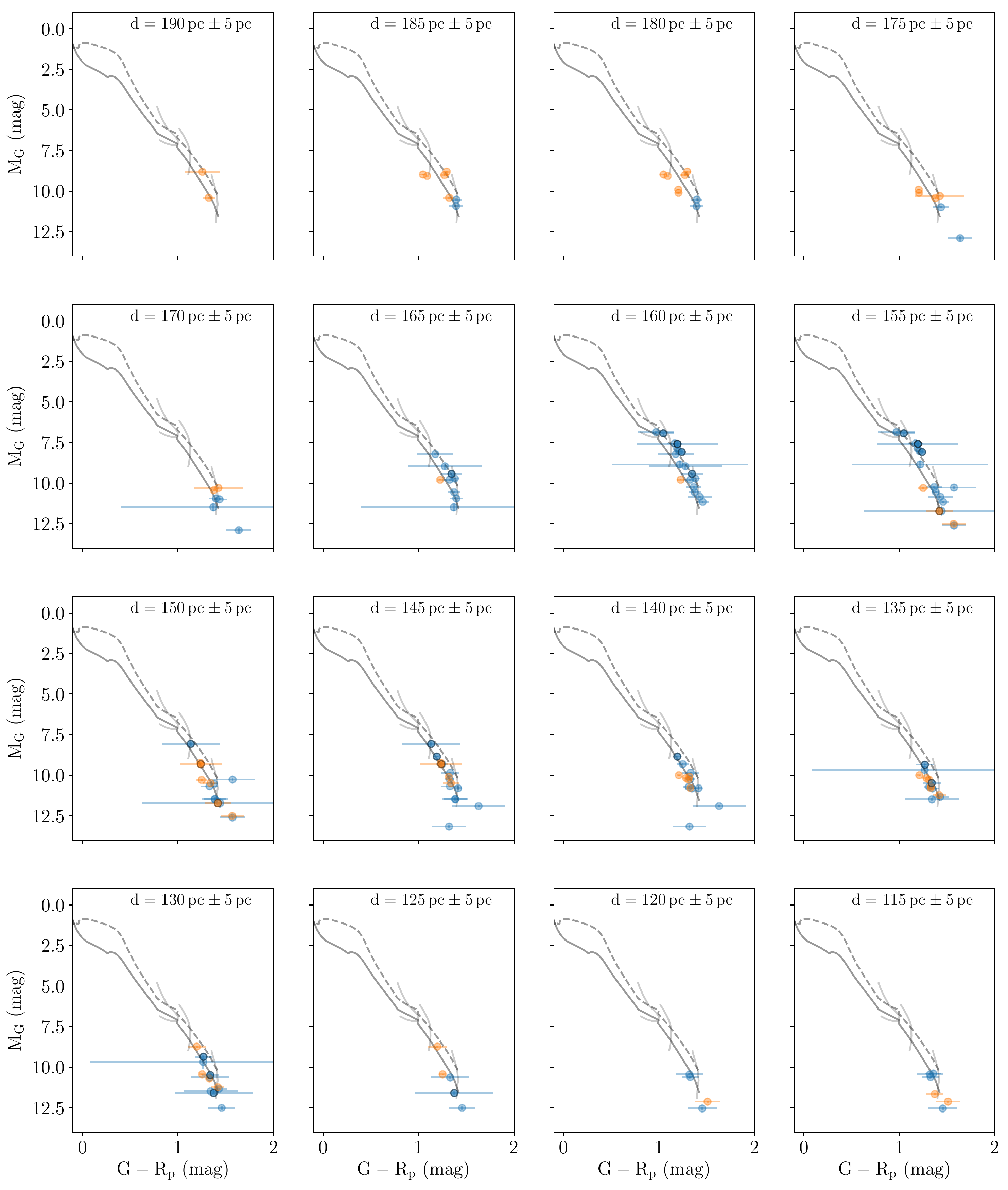}
\caption{Absolute magnitude vs. color for the new disk-bearing YSOs shown in Figures \ref{fig:pm_panel} and \ref{fig:pm_panel_ucl}. The YSO symbols are the same as those shown in Figure \ref{fig:pm}. The extinction-free isochrones correspond to 4\,Myr (dashed line, dark gray) and 12\,Myr (solid line, dark gray) \citep{marigo17}, with mass tracks of 0.1\msun, 0.4\msun, and 0.7\msun\ for reference.}
\label{fig:cmd_Lup_gaia}
\end{figure*}

\noindent The new disk-bearing YSOs can be split into probable US and V1062\,Sco members according to their spatial distribution (as delimited in Figure \ref{fig:s_dist_panel}). We separate these sources out to better analyze the new disked sources that overlap in proper motion and spatial distribution with the Lupus and UCL members. The three YSO groups were placed on Gaia color-magnitude diagrams (CMD), for the same distance slices as used for Figures \ref{fig:pm_panel} and \ref{fig:pm_panel_ucl}. The Lupus/UCL CMDs are displayed in Figure \ref{fig:cmd_Lup_gaia}, the US CMDs are shown in Figure \ref{fig:cmd_US_gaia}, and the V1062\,Sco CMDs are shown in Figure \ref{fig:cmd_v1062sco_gaia}.
The majority of the new YSOs with disks are very low mass stars, \hbox{M$_\star <$ 0.5\,\msun}, and they fall within an age range of 4\,Myr-12\,Myr. The youngest sources are located at 155\,pc-160\,pc and correspond to sources located near Lupus\,1, Lupus\,3, and Lupus\,4. The diskless variable sources are comparatively more massive than the disk-bearing new YSOs, but span the same age range. Sources older than 12\,Myr tend to be sources with anemic disks, and are mostly located at distances greater than 175\,pc.

\subsection{YSO evolutionary stage and mass distribution}

The majority of the new disk-bearing YSOs are part of Lupus or UCL, and are sources with thick disks. Table \ref{tab:newYSOs} breaks down the number of YSOs found per stellar group, divided into TDs and ADs following Table \ref{tab:alpha_wise}; it also lists sources that are variable (TDVs and ADVs). 

\begin{table}[!h]
    \caption{Summary of the number of new disk-bearing YSOs in the various stellar groups, in the area surveyed in this work. ``V'' denotes variable sources.}
    \label{tab:newYSOs}
    \centering
    \begin{tabular}{l|ccc|c}
    \cline{2-5}
      & Lupus/UCL & US & V1062\,Sco & Total (V)\\
    \hline
    \hline
    TD (TDV) & 34 (8) & 12 (1) & 10 (0) & 56 (9)\\
    AD (ADV) & 26 (2) & 7 (0) & 9 (0) & 42 (2) \\
    \hline
    Total (V) &  60 (10) &  19 (1) & 19 (0) & 98 (11)\\
    \hline
    \end{tabular}
\end{table}

\begin{figure}[!h]
\includegraphics[width=0.5\textwidth]{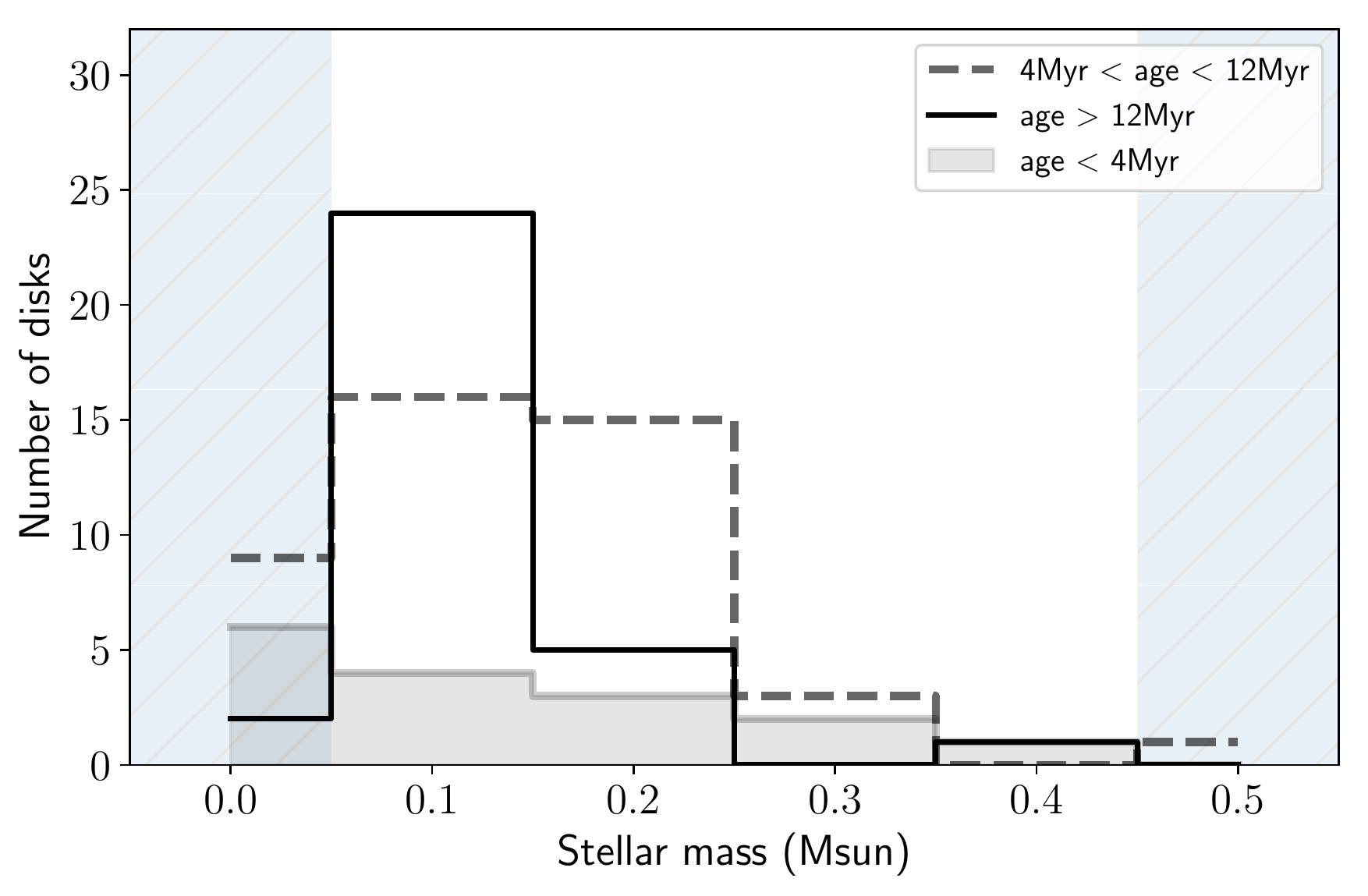}
\caption{Stellar mass distribution of the new disk-bearing YSOs in Lupus, for different age ranges. The first and last bins (shaded and hatched) are lower limits.}
\label{fig:Hdisks}
\end{figure}

\noindent We used the mass tracks in the Gaia CMDs (Figures \ref{fig:cmd_Lup_gaia}, \ref{fig:cmd_US_gaia}, and \ref{fig:cmd_v1062sco_gaia}) to derive stellar masses for the new disked YSOs. The sources were further divided into three groups according to their isochronal age: (i) age $<$ 4\,Myr, (ii) 12\,Myr $>$ age $>$ 4\,Myr, and (iii) age $>$ 12\,Myr. Figure \ref{fig:Hdisks} shows the mass distribution for the disk-bearing YSOs, for each of these age groups. The plot indicates that disk lifetimes are longer for lower mass sources - the oldest disked sources are the least massive ones.
 This result is consistent with previous observational findings showing there is a disk evolution dependence with stellar mass and that disks evolve later/slower around lower mass stars \citep[e.g.,][]{lada06a, teixeira12, ribas15}. The stellar masses probed here are much lower than previous studies as some of the YSOs are potential brown dwarfs, which has interesting implications for planet formation timescales around substellar objects. 

\subsection{Distance and age distribution}

\begin{table}[!h]
    \caption{Mean proper motions, according to age, for the new disk-bearing YSOs and members of Lupus and UCL. }
    \label{tab:pm-lup-ucl}
    \centering
    \begin{tabular}{c|ccc}
    \cline{2-4}
        & new YSOs with disks & Lupus & UCL \\
    \hline
    \hline
    \multicolumn{4}{c}{age $<$ 4\,Myr} \\
    \hline
      $<\mu_\alpha^*>$ & -11.5$\pm$1.6 & -10.2$\pm$1.0 & -19.3$\pm$3.7 \\
      $<\mu_\delta>$  & -23.2$\pm$0.4 & -23.4$\pm$0.8 & -22.2$\pm$2.5\\
    \hline
    \multicolumn{4}{c}{4\,Myr $<$ age $<$ 12\,Myr} \\
    \hline
      $<\mu_\alpha^*>$ & -16.1$\pm$3.5 & -10.9$\pm$1.3 & -17.7$\pm$2.9\\
      $<\mu_\delta>$  & -25.0$\pm$2.8 & 23.6$\pm$0.8 & -26.0$\pm$2.7\\
    \hline
    \multicolumn{4}{c}{age $<$ 12\,Myr} \\
    \hline
      $<\mu_\alpha^*>$ & -17.7$\pm$2.9 & -13.4$\pm$3.8 & -21.1$\pm$2.6\\
      $<\mu_\delta>$  & -26.0$\pm$2.7 & -23.5$\pm$0.6 & -27.9$\pm$3.1\\
     \hline
    \end{tabular}
\end{table}{}

\begin{figure*}[!h]
\centering
\includegraphics[width=0.8\textwidth]{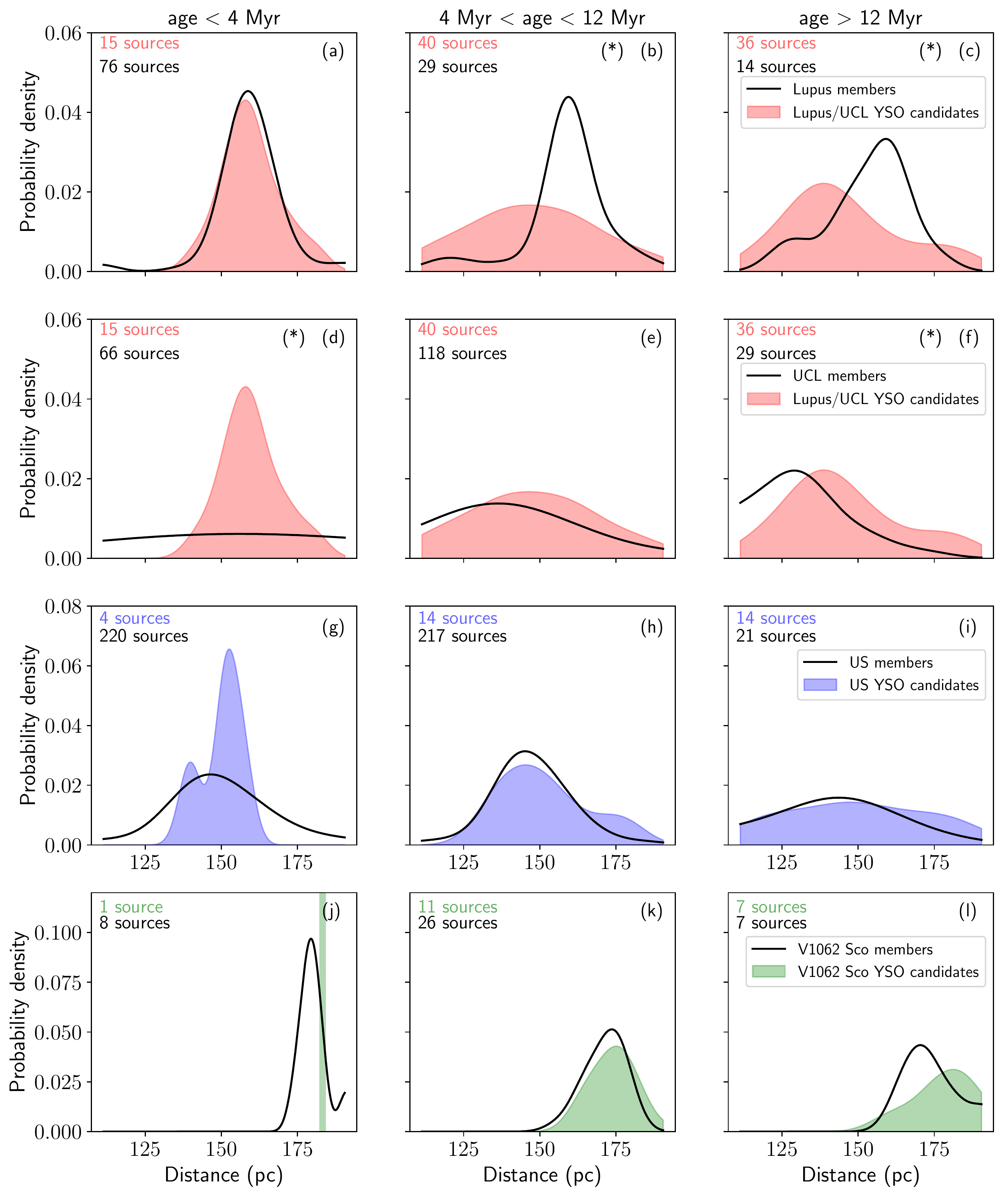}
\caption{Distance probability density function (PDF) of the new disk-bearing YSOs (color) and the previously known members (black) for Lupus, UCL, US, and V1062 Sco. Each panel indicates the number of sources for the respective PDF. Panels with (*) indicate that the null-hypothesis was rejected in at least one of the non-parametric statistic tests performed.}
\label{fig:age_dist}
\end{figure*}

\noindent Figure \ref{fig:age_dist} shows the distance Probability Density Functions (PDFs) of the new disk-bearing YSOs for three isochronal age groups aforementioned, compared to those of previously known members. They were built using a gaussian kernel distribution estimation with a bandwidth of 0.5. Panels (a) though (f) show PDFs (in red) of the majority of the new disk-bearing YSOs, except for those located in the US and V1062\,Sco regions. These PDFs are compared to the PDFs (in black) of the previously known members of Lupus in panels (a) through (c), and to those of UCL in panels (d) through (f). We performed nonparametric statistical hypothesis tests to determine if the paired PDFs were drawn from the same population (null hypothesis); specifically, the k-sample Anderson-Darling test, the Kolmogorov–Smirnov test, and the Wilcoxon-Mann-Whitney U test. The corresponding p-values are summarized in Table \ref{tab:p-values}.

\noindent The group of YSOs with isochronal age less than 4\,Myr is very likely part of Lupus.
This is ascertained by comparing panels (a) and (d), and because the statistical tests rejects the null hypothesis for panel (d). The group of YSOs with isochronal age greater than 4\,Myr and less than 12\,Myr is more likely to be part of UCL, since the null hypothesis is rejected for panel (b). The oldest group of YSOs, with isochronal age greater than 12\,Myr, do not appear to belong to Lupus nor to UCL because the null hypothesis is rejected for the PDFs in panels (c) and (f). This could possibly be explained if the group of either PDF being compared has contaminants, (i.e., is a mixture of populations). When comparing the average proper motions for the new disk-bearing YSOs with those of Lupus and UCL, for the same isochronal age greater than 12\,Myr, the YSOs are moving more similarly to UCL (see Table \ref{tab:pm-lup-ucl}). 

\noindent The bottom row of Figure \ref{fig:age_dist} shows the V1062\,Sco new disk-bearing YSOs compared to known members of this comoving group. The row corresponding to panels (g) through (i) shows the comparison of the new disk-bearing YSOs in US and its confirmed members. The new V1062\,Sco and US YSOs have distance PDFs consistent with those of previously known members.

\subsection{Gaia DR2 RUWE as an indicator of possible binarity}
\label{subsec:ruwe}

\begin{figure}[!h]
    \centering
    \includegraphics[width=0.49\textwidth]{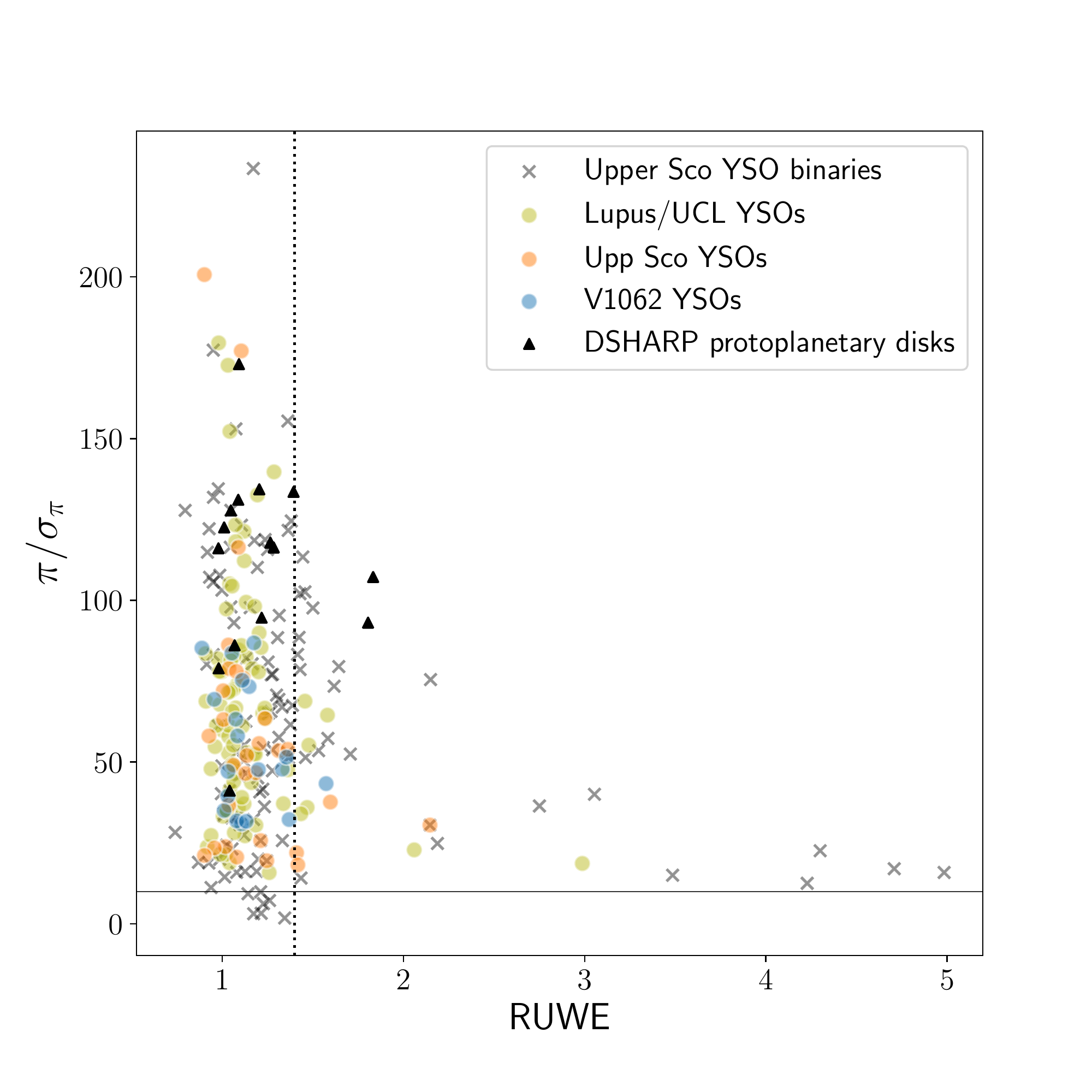}
    \caption{Parallax over error ($\pi / \sigma_\pi$) ve. RUWE of the three new groups of YSOs identified in this work, compared to sources with resolved protoplanetary disks \citep[DSHARP,][]{andrews18} and known binaries in Upper Sco \citep{tokovinin20}}
    \label{fig:ruwe}
\end{figure}

The re-normalized unit weight error, RUWE, is an optimized parameter that is used to assess the goodness of the astrometric solution of Gaia DR2 data for single sources \citep{lindegren18}. Unresolved binary systems, on the other hand, can cause the astrometric model to perform poorly due to the wobble of the Gaia photocenter. Marginally resolved sources and variable sources may also present excess RUWE \citep{belokurov20}.
Figure \ref{fig:ruwe} shows the parallax over the error, $\pi / \sigma_\pi$, versus RUWE of the new YSO sources identified in this work, namely in Lupus/UCL, Upper Sco, and V1062 Sco. The great majority of the sources have values of RUWE less than 1.4, indicating a robust astrometric solution. For comparison, we also plot sources with resolved disks from the DSHARP project \citep{andrews18} and spectroscopically confirmed binaries in Upper Sco from \citet[][ binary separations $<$0.7\arcsec]{tokovinin20}. 
As can be seen, sources with disks may have larger values of RUWE, and YSO binaries can indeed have much larger values of RUWE. In order to not remove interesting sources from our new YSO samples we have therefore opted to retain sources with values of RUWE larger than the canonical value of 1.4. 

\section{On the origin of the new YSOs with disks}
\label{sec:discussion}

The previous section has shown that the new disk-bearing YSOs that are candidate members of the US and V1062\,Sco comoving groups are relatively well characterized with respect to proper motion, on-sky spatial distribution, distance distribution, and age. Tables \ref{tab:new_US_ysos} and \ref{tab:new_V1062sco_ysos} list the source designations and disk type for US and V1062\,Sco, respectively.

\noindent The remaining YSOs in our sample (Table \ref{tab:new_Lup_ysos}) are likely composed of sources from both Lupus and UCL, as they present properties that are common to both of these populations; the emerging picture is that the youngest of these YSOs are likely Lupus members, while the older sources are likely part of UCL.
Comparison of our new disk-bearing sources in Table \ref{tab:new_Lup_ysos} with the YSOs identified by \citet{damiani19} yielded 37 out of 60 matches: 45\% of the sources were found to belong to the UCL diffuse group D1, and 35\% to the  UCL diffuse group D2 (further divided into 27\% from the D2a sub-group and 8\% from the D2b sub-group).
Of our new disks in US, 12/19 were identified as belonging to either the diffuse population in Upper Sco USC-D2 (32\%), to UCL-D1 (32\%), or to UCL-D2a (5\%). Regarding the new disks in V1062\,Sco, 14/19 were also previously identified by \citet{damiani19} as candidate members of the V1062\,Sco/UCL-1 compact group.

\noindent We discuss in this section the region morphology and kinematics to explore the origin and connection between Lupus, UCL, US, and V1062\,Sco.

\subsection{Evolution of the 3D spatial distribution}
\begin{figure*}[!h]
    \centering
    \includegraphics[width=0.92\textwidth]{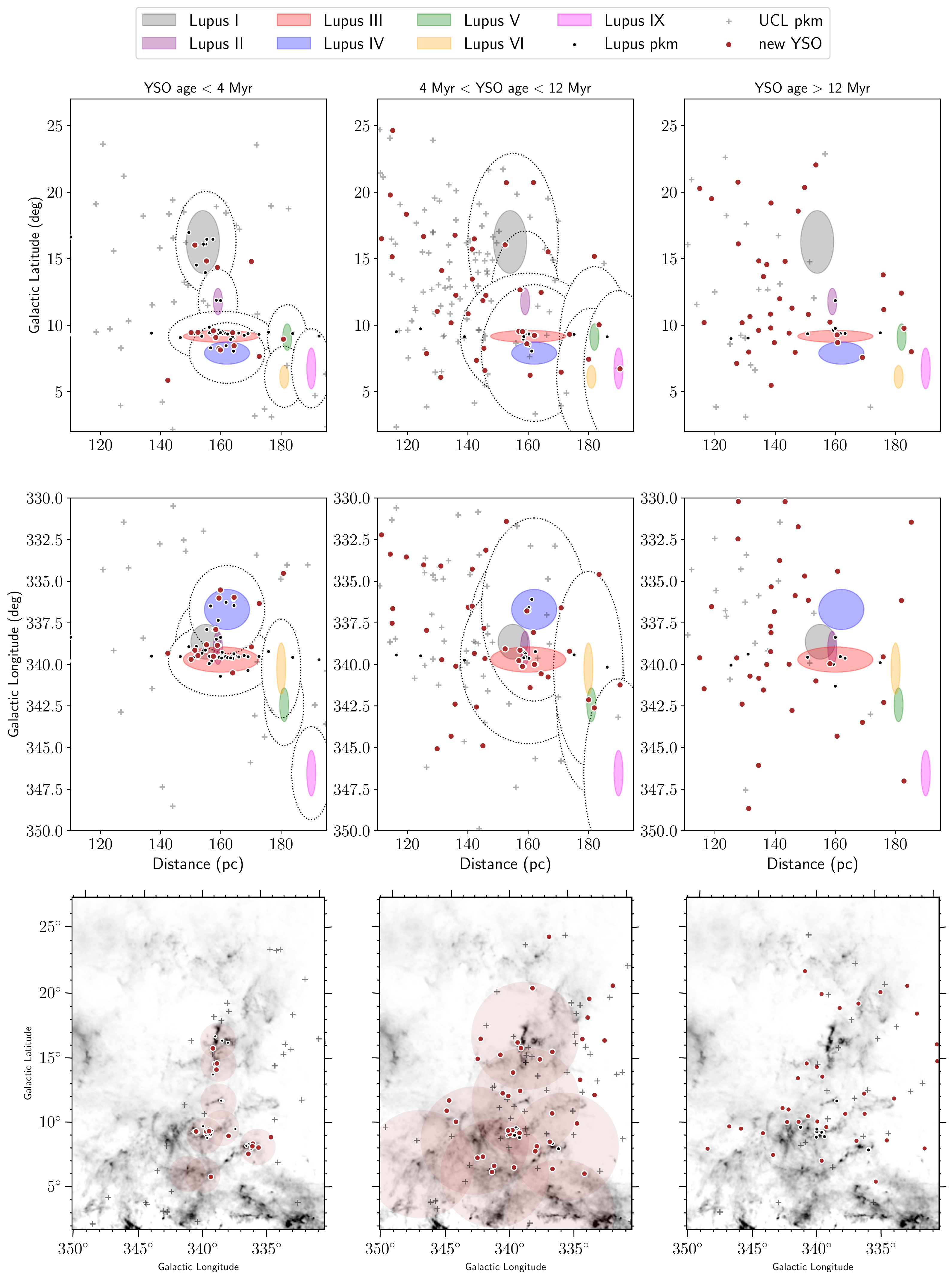}
    \caption{Top and middle panels: spatial distribution along the line-of-sight of the new disk-bearing YSOs and previously known members (pkm) in Lupus or UCL, for three age groups (less than 4\,Myr, between 4\,Myr and 12\,Myr, and greater than 12\,Myr). Seven Lupus clouds are  represented schematically by the filled colored ellipses. The 4\,Myr travel radii is represented by dotted ellipses, centered on each cloud and is shown in the left columns; the middle columns show the same but for 12\,Myr travel radii. 
    Bottom panels: spatial distribution of new Lupus or UCL YSOs, and previous known members, superimposed on the Planck A$\_\mathrm{v}$ map (the colorscale is the same as that shown in Figure \ref{fig:clouds}). The shaded red circles represent the 4\,Myr and 12\,Myr travel radii, for the left and middle panels, respectively. }
    \label{fig:3D_age}
\end{figure*}

\noindent We attempt to reconstruct the 3D distribution of the new disk-bearing YSOs and the Lupus clouds to better determine how the Lupus and UCL stellar groups are related. 
Figure \ref{fig:3D_age} shows in the upper (middle) row the Galactic Latitude (Longitude) as a function of distance for these sources. Each row has three panels, corresponding to the three iscohronal age groups.
These panels also include a schematic representation of the Lupus clouds (the central Galactic coordinates were taken from \citet{hara99}
and the width of the filled ellipses are merely representative and not to be taken to scale). The distances to Lupus\,1, Lupus\,2, Lupus\,3, and Lupus\,4 were derived from the distances to their previously known members \citep{dzib18}, and for clouds with no known associated YSOs we use the distances calculated by \citet{zucker19} (we exclude the clouds Lupus\,7 and Lupus\,8, due to the lack of distance estimates). To constrain the origin of a YSO, we explore its travel radius, specifically, the projected distance traversed by a source since its birth assuming a 1D velocity of 1\kms. We represent the area corresponding to the travel radius as dotted-line ellipses, assuming the birth-sites are at the center of the clouds: The leftmost column shows ellipses corresponding to the travel radii of 4\,Myr for each cloud, while the middle column to that of 12\,Myr. Star-formation occurs in dense filaments that thread the cloud, and so the ellipses show the least amount a source would travel and are merely indicative. In addition to that, we do not have strong constraints on the depth of the clouds as mentioned above.
The bottom row shows the distribution of these sources on the plane-of-the-sky, overplotted on a Planck extinction map; the shaded red areas mark the 4\,Myr and 12\,Myr travel radii for the left and middle panels, respectively.

\noindent The panels from the left column shows that the new disk-bearing YSOs, with isochronal ages  less 4\,Myr, are mostly confined to the clouds Lupus\,1, Lupus\,3, and Lupus\,4. Apart from three sources that appear to be located nearer or farther, these YSOs are all within the travel radii from the aforementioned clouds and so clearly associated with the Lupus molecular cloud complex (as already indicated by panels (a) and (d) of Figure \ref{fig:age_dist}). The three exceptions all have proper motions well within the 5-$\sigma$ of the average proper motions of previously known Lupus members.

\noindent The panels from the middle column show the distribution of sources with isochronal ages between 4\,Myr and 12\,Myr. The area traversed by a 12\,Myr travel radius is again represented by dotted-line ellipses. The new YSO sources are much more dispersed comparatively to the panels in the left column. Sources within the ellipses could still have formed within the Lupus clouds; Lupus\,3 has a distinct group of sources centered on the cloud, while the other clouds in the complex have a few sources that appear associated with them. There is a larger group of sources that are located outside the dotted-line ellipses and the shaded area on the bottom panel, and as such are unlikely to have formed within the Lupus cloud. These sources show a distribution similar to the previously known UCL members and are likely UCL members themselves. The new disk-bearing YSOs ages between 4\,Myr and 12\,Myr are thus a mixture of probable members from Lupus and UCL. 

\noindent The panels from the right column show sources with isochronal ages greater than 12\,Myr. The Lupus clouds do not appear to have any clustering of new disk-bearing YSOs, although Lupus\,3 has a clustering of previously known members. This may place an age constraint on the Lupus\,3 cloud: its lifetime is at least 12\,Myr long. The new disk-bearing YSOs in this age bracket show the same distribution as UCL members. As can be seen, the spatially dispersed disk population is older than the disks associated with the molecular clouds, in agreement with the results from \citet{damiani19}.\\

\noindent Regarding the ages of the clouds, Lupus\,1 is the youngest star forming cloud \citep{krause18}, followed by the Lupus\,3 and Lupus\,4 clouds. Lupus\,3 in particular appears to have formed sources over a wider period of time compared to the other clouds \citep[in agreement with ][who also find that there is a large spread in apparent ages of the youngest compact groups]{damiani19}.
The  Lupus\,5 and Lupus\,6 clouds appear to be older which is consistent with the recent findings of \citet{melton20}. We do not have enough information to constrain the relative ages of the remaining clouds in the complex. Within the Lupus complex, star formation would thus seem to be progressing from the more distant and lower clouds, to the closer and higher latitude clouds.

\subsection{Kinematics of Lupus complex}

\begin{figure*}[!h]
\sidecaption
\includegraphics[width=12cm]{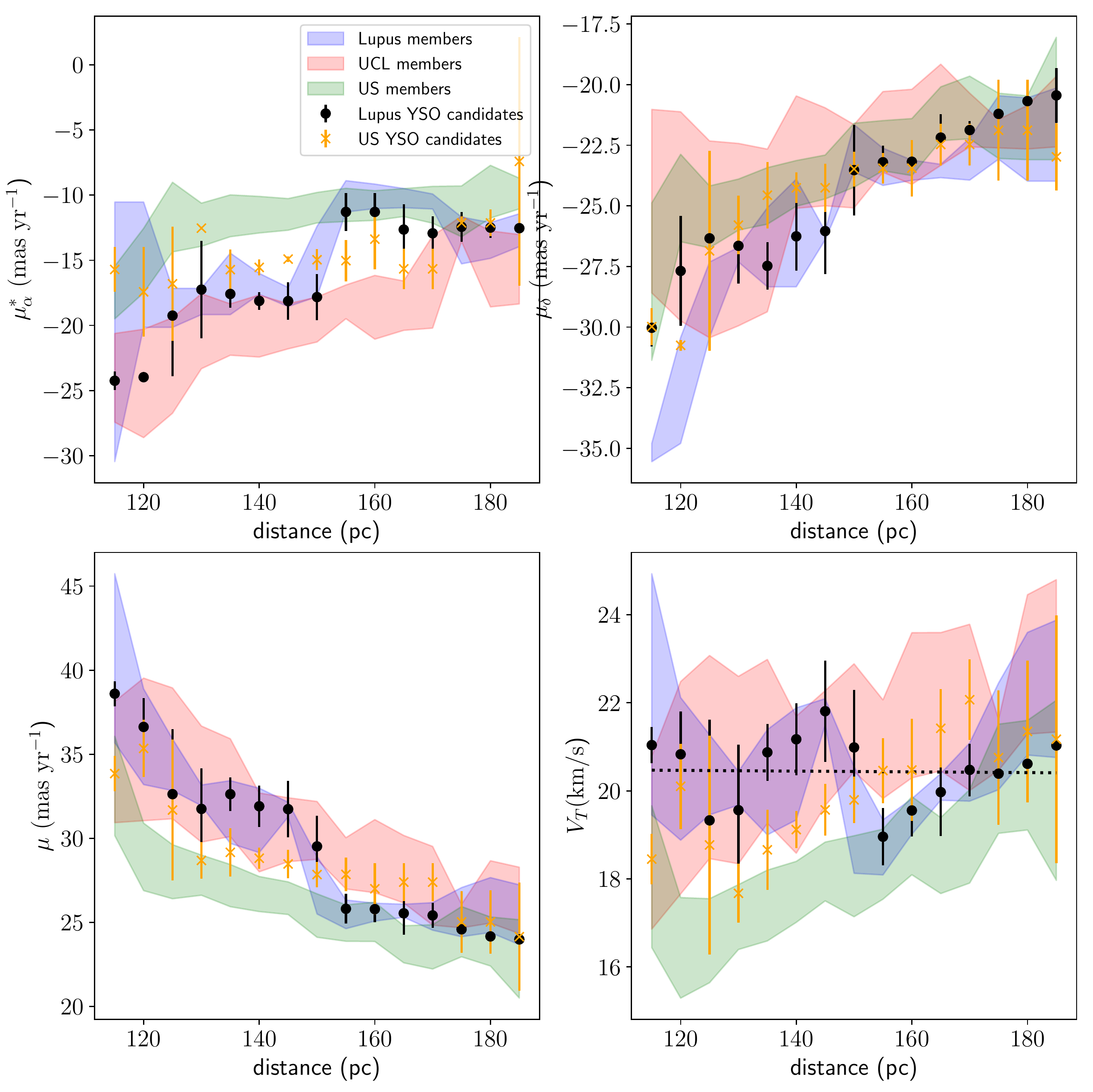}
\caption{Overall proper motion and transverse velocity as a function of distance for the previously known members and new disk-bearing YSOs of the Lupus, UCL, and US groups.
The data points denote the median values for each distance slice and the error bars corresponds to the respective median absolute deviation; the black data points correspond to values for the new disk-bearing YSOs in Lupus or UCL, while the orange data points correspond to values for the new disk-bearing YSOs in US. Each panel shows the corresponding median absolute deviation for known members for Lupus, US, and UCL (shaded bands). The dotted line in the bottom right panel corresponds to a linear fit to the transverse velocity over distance for the new YSO Lupus/UCL members: 
{\hbox{$ {V_\mathrm{T} = 20 \pm 1}$\,\kms}. }}
\label{fig:vt}
\end{figure*}

We analyzed the kinematics of the US, Lupus, and UCL groups as a function of distance. Figure \ref{fig:vt} shows four panels with the median values for each distance slice of the proper motions along R.A. and Dec, $\mu^*_\alpha$ and $\mu_\delta$; the total proper motion, $\mu = \sqrt{\mu_\alpha{^*}{^2} + \mu_\delta^2}$, and the transverse velocity, $V_\mathrm{T}$ which is given by:

\begin{equation}
V_\mathrm{T} = \frac{4.74}{\pi} * \mu\ \ \ \mathrm{km\,s^{-1}.}
\label{eq:vt}
\end{equation}

\noindent The three groups of previously known members are represented by a curve or band where the width is the median absolute deviation; previously known Lupus members correspond to the blue shaded band, UCL by a red band, and US by a green band. The new disk-bearing YSOs in US and Lupus/UCL are also shown as data points with error bars in orange and black, respectively. The new US YSOs show the same functional variation with distance as the previously know US members, albeit with higher velocities.
These plots also show that the new disk-bearing candidate members of Lupus/UCL follow the kinematics of the previously known Lupus members, however, they are also consistent with the motion of UCL members, particularly for distances smaller than 150\,pc. 
Their transverse velocity can be described as being constant
\footnote{A linear regression fit to the data points corresponding to the new Lupus/UCL disk-bearing sources (bottom right panel of Figure \ref{fig:vt}) yields \hbox{$ V_\mathrm{T}(\kms) = (20 \pm 1) + (0.000 \pm 0.001) \cdot \mathrm{Distance(pc).}$}}, with \hbox{$V_\mathrm{T} = 20 \pm 1$\,\kms}, meaning the new disk-bearing sources that are candidate members of Lupus/UCL are moving at the same speed irrespective of their distance. There is some sub-structure in the velocity around 155\,pc, which is the distance where the youngest sources are located, but otherwise the transverse velocity of the new disk-bearing cannot be differentiated from those of Lupus or UCL. The emerging picture is that Lupus sources are part of UCL - the previously known Lupus sources and the new disk-bearing YSOs with ages less than 4\,Myr are the tail end of star formation in UCL, in much the same way that $\rho$\,Ophiuchus is the tail end of star formation of US.

\section{Conclusion}
\label{sec:conclusion}


The main results of our survey are:
\begin{enumerate}
\item Using Gaia DR2 and ALLWISE data, we identify 98 new low-mass YSOs within the wider Lupus complex. These sources were identified via their SED slope in the mid-infrared, and variability in the mid-infrared and optical. The evolutionary state of the new disk-bearing YSOs are as follows: 56 are sources with thick disks and 42 are sources with anemic disks.
\item We use the Upper Scorpius \emph{(l,b)} boundary to identify potential new US members, likewise for the V\,1062\,Scorpii sub-group. The proper motion of these new members fall within 5$\sigma$ of the proper motion of previously confirmed members. The new disk-bearing YSOs are thus distributed:
19 new disks in the Upper Scorpius region, 19 new disks in region of the V\,1062\,Scorpius sub-group, and 60 new disks in the region that is occupied by both Lupus and UCL.
\item The new disk-bearing YSOs are low mass sources, most with masses less than 0.4\msun. Dividing the YSOs in three age groups shows that the stellar mass of disk bearing sources decreases with age.
\item A combined analysis of the distance, age, and kinematic distribution of the new disk-bearing YSOs that fall within the Lupus/UCL region indicate that Lupus and UCL are not distinct groups because many of their properties are indistinguishable. On the contrary, the emerging picture is that the latest star formation episode taking place in the Lupus clouds (particularly Lupus\,1, Lupus\,3, and Lupus\,4) is the tail end of star formation of UCL.

\end{enumerate}

\begin{acknowledgements}
We would like to thank the anonymous referee for their valuable comments that helped improve the manuscript.
This project was supported by STFC grant ST/R000824/1. This  work  has  made  use  of data  from  the  European  Space  Agency  (ESA)  mission Gaia (https://www.cosmos.esa.int/gaia),  processed  by  the Gaia Data  Processing  and  Analysis Consortium (DPAC, https://www.cosmos.esa.int/web/gaia/dpac/consortium). Funding for the DPAC has been provided by national institutions, in particular the institutions participating in the Gaia Multilateral Agreement. 
This publication makes use of data products from the Wide-field Infrared Survey Explorer, which is a joint project of the University of California, Los Angeles, and the Jet Propulsion Laboratory/California Institute of Technology, and NEOWISE, which is a project of the Jet Propulsion Laboratory/California Institute of Technology. WISE and NEOWISE are funded by the National Aeronautics and Space Administration.
This research has made use of the VizieR catalog access tool, CDS, Strasbourg, France. 
This research has made use of the NASA/ IPAC Infrared Science Archive, which is operated by the Jet Propulsion Laboratory, California Institute of Technology, under contract with the National Aeronautics and Space Administration.
This research has made use of Python, https://www.python.org, NumPy \citep{vanderwalt11}, and Matplotlib \citep{hunter07}.
This research made use of APLpy, an open-source plotting package for Python and hosted at  http://aplpy.github.com \citep{robitaille12}. This research made use of Astropy, a community-developed core Python package for Astronomy \citep{robitaille13}. This  research  made  use  of  TOPCAT,  an  interactive  graphical viewer  and  editor  for  tabular  data  \citep{taylor05}.  

\end{acknowledgements}

\onecolumn

\renewcommand{\thefootnote}{\fnsymbol{footnote}}
\begin{longtable}{cccc}
\caption{\label{tab:new_Lup_ysos} List of new  disk-bearing YSO candidate members of Lupus and/or UCL.}\\
\hline\hline
Gaia designation & ALLWISE designation  &
Disk type & notes\protect\footnotemark[1] \\
\hline
\endfirsthead
\caption{continued.}\\
\hline
Gaia designation & ALLWISE designation & 
Disk type & notes\protect\footnotemark[1] \\
\hline\hline
\endhead
\hline
\endfoot
\hline
\multicolumn{4}{c}{New candidate members with disks}\\
\hline
Gaia DR2 5995206142536938112 & WISEA J155806.16-420837.2 & Thick &  Ovar \\ 
Gaia DR2 5995154907858605312 & WISEA J155859.23-421613.2 & Thick &  Ovar MIRvar \\ 
Gaia DR2 5992969838966885632 & WISEA J162044.68-414601.1 & Thick &  \\ 
Gaia DR2 5993786226354952960 & WISEA J161909.13-410121.1 & Thick &  MIRvar \\ 
Gaia DR2 6023700742290627712 & WISEA J163024.81-342048.3$^\dagger$ & Thick &  \\ 
Gaia DR2 5996898290938990720 & WISEA J160330.70-402434.1 & Thick &  MIRvar \\ 
Gaia DR2 6018623025796836224 & WISEA J162913.21-373621.1$^\dagger$ & Thick &  \\ 
Gaia DR2 6017201425972688512 & WISEA J162654.39-400152.4$^\dagger$ & Thick &  \\ 
Gaia DR2 6027035488023161472 & WISEA J164428.24-332948.2$^\dagger$ & Thick &  LR \\ 
Gaia DR2 5988278360303787392 & WISEA J155850.50-450949.7$^\dagger$ & Thick &  \\ 
Gaia DR2 6018187752309833472 & WISEA J162650.40-382518.9 & Thick &  \\ 
Gaia DR2 6010730074180913024 & WISEA J160148.11-360847.0 & Thick &  \\ 
Gaia DR2 6011576663773784832 & WISEA J154826.33-352544.3$^\dagger$ & Thick &  \\ 
Gaia DR2 6203852561384421632 & WISEA J150300.12-344728.6 & Thick &  \\ 
Gaia DR2 6023203655639434752 & WISEA J162208.06-340510.6$^\dagger$ & Thick &  \\ 
Gaia DR2 6012073922207639424 & WISEA J155501.54-341228.8$^\dagger$ & Thick &  MIRvar \\
Gaia DR2 6011165313293141760 & WISEA J160118.68-343733.4$^\dagger$ & Thick &  \\ 
Gaia DR2 6012155767105823616 & WISEA J155235.72-334429.0$^\dagger$ & Thick &  LR \\ 
Gaia DR2 5998419156031525120 & WISEA J161245.39-371135.8$^\dagger$ & Thick &  \\ 
Gaia DR2 6005508630897881856 & WISEA J151009.63-391830.0 & Thick &  \\ 
Gaia DR2 6022369680375649792 & WISEA J162239.55-351306.2 & Thick &  MIRvar \\ 
Gaia DR2 6003224674675574272 & WISEA J153439.71-391644.7$^\dagger$ & Thick &  \\ 
Gaia DR2 6206733728527881984 & WISEA J153206.47-323024.9$^\dagger$ & Thick &  \\ 
Gaia DR2 6206819456073299072 & WISEA J152532.95-330751.8 & Thick &  \\ 
Gaia DR2 6007156047211408384 & WISEA J151724.13-375637.6$^\dagger$ & Thick &  \\ 
Gaia DR2 6207172635526215936 & WISEA J151627.26-332609.4 & Thick &  \\ 
Gaia DR2 6022863842143473920 & WISEA J161333.63-353121.8 & Thick &  \\ 
Gaia DR2 6036287126758348160 & WISEA J160004.22-324755.6 & Thick &  \\ 
Gaia DR2 5995782836390205696 & WISEA J154702.13-414809.6$^\dagger$ & Thick &  LR \\ 
Gaia DR2 6023457063014473728 & WISEA J161842.78-333843.3 & Thick &  \\ 
Gaia DR2 6199998665752927744 & WISEA J151734.49-354229.5 & Thick &  \\ 
Gaia DR2 5996365543191043712 & WISEA J155213.43-395608.5 & Thick &  MIRvar \\ 
Gaia DR2 6201266063359289472 & WISEA J150657.36-341438.3 & Thick &  MIRvar \\ 
Gaia DR2 6207841245973096448 & WISEA J152741.07-311718.6 & Thick &  \\ 
Gaia DR2 5994961088885240192 & WISEA J160225.93-423004.4$^\dagger$ & Anemic &  \\ 
Gaia DR2 6017949776765321216 & WISEA J162429.70-394853.5$^\dagger$ & Anemic &  \\ 
Gaia DR2 5995500571145275520 & WISEA J155025.81-424107.6 & Anemic &  LR \\ 
Gaia DR2 6022093492495726976 & WISEA J162552.37-355039.1$^\dagger$ & Anemic &  \\ 
Gaia DR2 5988956243573459968 & WISEA J154024.20-451846.6$^\dagger$ & Anemic &  \\ 
Gaia DR2 5991824560190531712 & WISEA J160748.45-431236.3$^\dagger$ & Anemic &  \\ 
Gaia DR2 5995299566669151872 & WISEA J160041.04-412446.3$^\dagger$ & Anemic &  \\ 
Gaia DR2 5996621729410099712 & WISEA J160606.36-410344.6$^\dagger$ & Anemic &  \\ 
Gaia DR2 6018168893133793536 & WISEA J162501.52-384017.9 & Anemic &  \\ 
Gaia DR2 5994020834646630400 & WISEA J161702.26-404010.1$^\dagger$ & Anemic &  LR \\ 
Gaia DR2 6023845637312790656 & WISEA J163252.53-333319.6$^\dagger$ & Anemic &  \\ 
Gaia DR2 6011432803852543744 & WISEA J155323.73-352744.2$^\dagger$ & Anemic &  \\ 
Gaia DR2 6011649338920614400 & WISEA J155259.68-345627.4$^\dagger$ & Anemic &  \\ 
Gaia DR2 6008346268558263936 & WISEA J154530.05-385904.1$^\dagger$ & Anemic &  \\ 
Gaia DR2 6011350787155212800 & WISEA J155349.77-354904.3 & Anemic &  \\ 
Gaia DR2 5998447635942264576 & WISEA J160828.27-372301.9$^\dagger$ & Anemic &  \\ 
Gaia DR2 6022569963302063744 & WISEA J161215.45-355553.9$^\dagger$ & Anemic &  \\ 
Gaia DR2 6002841735399039616 & WISEA J153348.52-405526.7 & Anemic &  MIRvar \\ 
Gaia DR2 6013156906798029696 & WISEA J153626.68-361056.1$^\dagger$ & Anemic &  \\ 
Gaia DR2 6007508371970923264 & WISEA J152407.48-364939.1$^\dagger$ & Anemic &  \\ 
Gaia DR2 6199706195661939456 & WISEA J150947.12-362703.5$^\dagger$ & Anemic &  \\ 
Gaia DR2 5996355235268966528 & WISEA J155041.21-401359.3$^\dagger$ & Anemic &  Ovar \\ 
Gaia DR2 6212274476855985024 & WISEA J151211.08-285226.5 & Anemic &  \\ 
Gaia DR2 5996511709515774592 & WISEA J155641.75-391345.0$^\dagger$ & Anemic &  \\ 
Gaia DR2 6210323393473690880 & WISEA J153350.86-283445.1 & Anemic &  \\ 
Gaia DR2 6208688488402623488 & WISEA J153413.93-304614.8 & Anemic &  \\ 
\hline
\multicolumn{4}{c}{Previously known members with disks}\\
\hline
Gaia DR2 5995298260998836736 & WISEA J160049.42-413004.1$^\dagger$ & Thick &  \\ 
Gaia DR2 5997493401600889856 & WISEA J161019.82-383607.0$^\dagger$ & Thick &  \\ 
Gaia DR2 5997459484227732864 & WISEA J160859.54-385627.7 & Thick &  \\ 
Gaia DR2 5997421658465250560 & WISEA J161013.04-384616.8 & Thick &  \\ 
Gaia DR2 5994747367001754240 & WISEA J155724.00-424004.8 & Thick &  MIRvar \\ 
Gaia DR2 5994761080849322368 & WISEA J155925.21-423506.9$^\dagger$ & Thick &  \\ 
Gaia DR2 5997071532738041216 & WISEA J160737.73-392138.9$^\dagger$ & Thick &  \\ 
Gaia DR2 5997416573223873536 & WISEA J161051.57-385313.9 & Thick &  \\ 
Gaia DR2 5997456224362720896 & WISEA J161159.78-382338.5 & Thick &  \\ 
Gaia DR2 5997464711218416384 & WISEA J160855.28-384848.3$^\dagger$ & Thick &  \\ 
Gaia DR2 5997087510016827776 & WISEA J160752.30-385806.3 & Thick &  \\ 
Gaia DR2 5997493397287745408 & WISEA J161018.55-383612.7$^\dagger$ & Thick &  \\ 
Gaia DR2 6011500389453302272 & WISEA J154930.72-354951.6 & Thick &  \\ 
Gaia DR2 6011573266459331072 & WISEA J154750.61-352835.6$^\dagger$ & Thick &  \\ 
Gaia DR2 6014693405579853184 & WISEA J154518.50-342124.7 & Thick &  LR \\ 
Gaia DR2 6014695913840754176 & WISEA J154457.87-342339.4 & Thick &  \\ 
Gaia DR2 6010483616079976448 & WISEA J155602.08-365528.5$^\dagger$ & Thick &  \\ 
Gaia DR2 6015000547281686784 & WISEA J154433.89-335254.3 & Thick &  LR \\ 
Gaia DR2 5997084452000075904 & WISEA J160804.75-390449.7$^\dagger$ & Anemic &  \\ 
Gaia DR2 5997845210966001536 & WISEA J160603.77-385952.3$^\dagger$ & Anemic &  \\ 

\footnotetext[0]{$\dagger$: possible source confusion}
\footnotetext[1]{Ovar: flagged as variable by Gaia DR2; MIRvar: flagged as variable by ALLWISE; LR: Gaia DR\,2 large RUWVE (defined as greater than 1.4)}
\end{longtable}

\twocolumn


\bibliographystyle{aa}

\onecolumn
\begin{appendix}

\section{Gaia and ALLWISE ADQL queries}
\label{sec:gaiaquery}

\noindent ADQL-2.0 script used to select sources from the ESA Gaia table \texttt{gaiadr2.gaia\_source}:

\begin{verbatim}
SELECT * 
FROM gaiadr2.gaia_source
WHERE l BETWEEN 330 AND 349 
    AND
    b BETWEEN 1.6 AND 27.6
    AND
    parallax BETWEEN 4 AND 12
    AND
    pmra_error < 1 AND pmdec_error <1
\end{verbatim}

\noindent ADQL-2.0 script used to cross-match the ESA Gaia tables \texttt{gaiadr2.gaia\_source} and \texttt {gaiadr2.allwise\_best\_neighbour}:

\begin{verbatim}
SELECT * 
FROM gaiadr2.gaia_source as gaia 
INNER JOIN gaiadr2.allwise_best_neighbour as allwise
ON gaia.source_id=allwise.source_id
WHERE gaia.l BETWEEN 330 AND 349 
    AND
    gaia.b BETWEEN 1.6 AND 27.6
    AND
    gaia.parallax BETWEEN 4 AND 12
    AND
    gaia.pmra_error < 1 AND gaia.pmdec_error <1
\end{verbatim}

\noindent ADQL-2.0 script to query the TAPVizieR table \texttt{IV/35/wn18\_b10} from \citet{wilson18} at \url{http://tapvizier.u-strasbg.fr/adql/?IV/35}:

\begin{verbatim}
SELECT 
"IV/35/wn18_b10".RAICRS,  "IV/35/wn18_b10".DEICRS, 
"IV/35/wn18_b10".GAIA,  "IV/35/wn18_b10".WISE, 
"IV/35/wn18_b10".RAWdeg,  "IV/35/wn18_b10".DEWdeg, 
"IV/35/wn18_b10".gmag, "IV/35/wn18_b10".W1mag, 
"IV/35/wn18_b10".W2mag,  "IV/35/wn18_b10".W3mag, 
"IV/35/wn18_b10".W4mag,  "IV/35/wn18_b10".MatchP, 
"IV/35/wn18_b10".eta,  "IV/35/wn18_b10".xi, 
"IV/35/wn18_b10".ContP1,"IV/35/wn18_b10".ContP10, 
"IV/35/wn18_b10".avgCont
FROM "IV/35/wn18_b10"
WHERE 
1=CONTAINS(POINT('ICRS',"IV/35/wn18_b10".RAICRS,
    "IV/35/wn18_b10".DEICRS), CIRCLE('GALACTIC', 
    340.25, +15.25, 18.))
\end{verbatim}

\clearpage

\section{Spatial distribution of the YSOs}

\begin{figure*}[!h]
\centering
\includegraphics[width=0.9\textwidth]{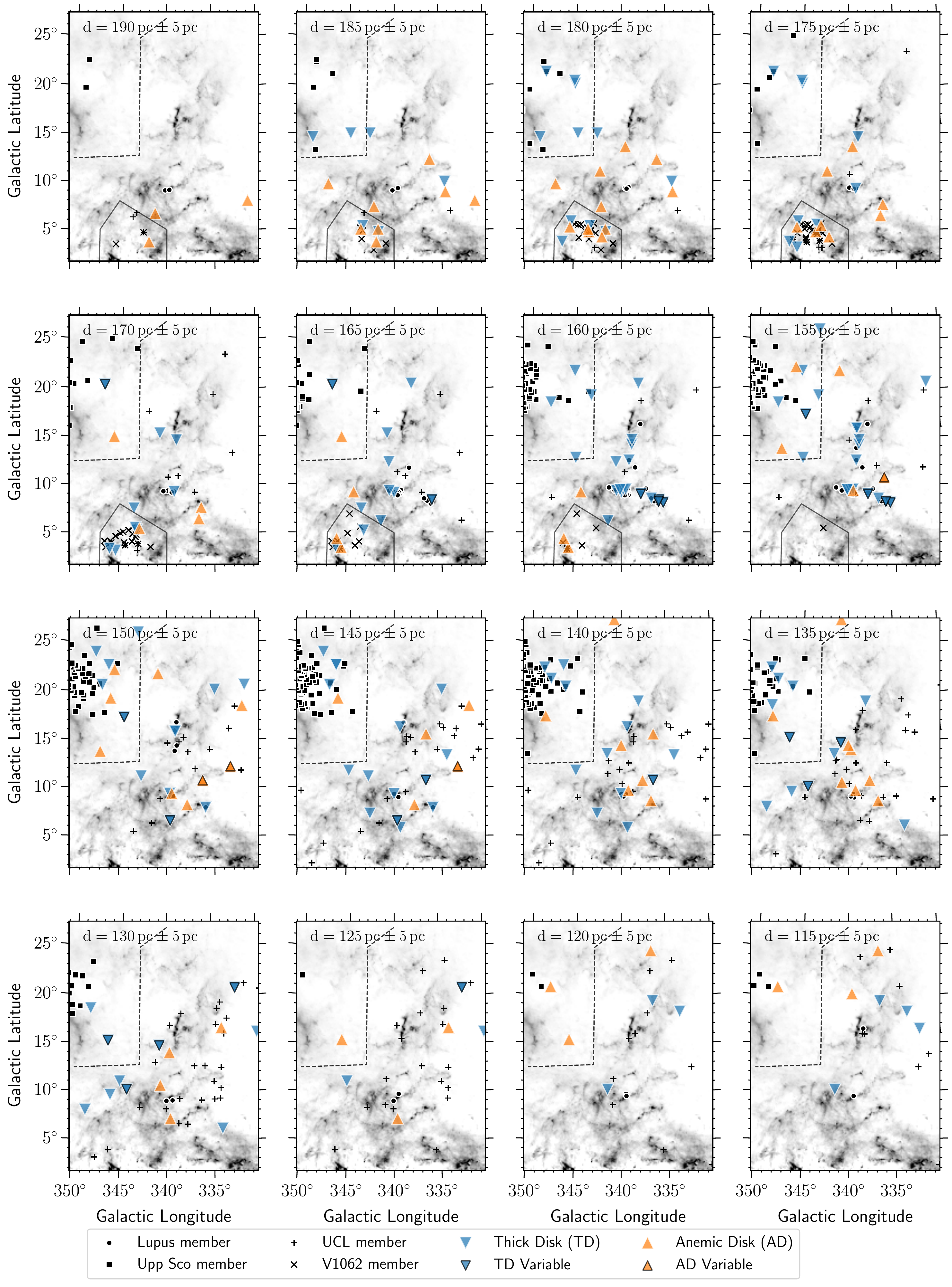}
\caption{Spatial distribution for the YSOs, and previously known members (pkm), for different distance slices, superimposed on the Planck \av\ map (colorscale is the same as in Figure \ref{fig:clouds}). The Upper\,Sco and V1062\,Sco regions are marked by dashed and solid lines, respectively.} 
\label{fig:s_dist_panel_appendix}
\end{figure*}

\clearpage

\section{Upper Scorpius}
\label{sec:uppsco}

\begin{figure*}[!h]
\centering
\includegraphics[width=\textwidth]{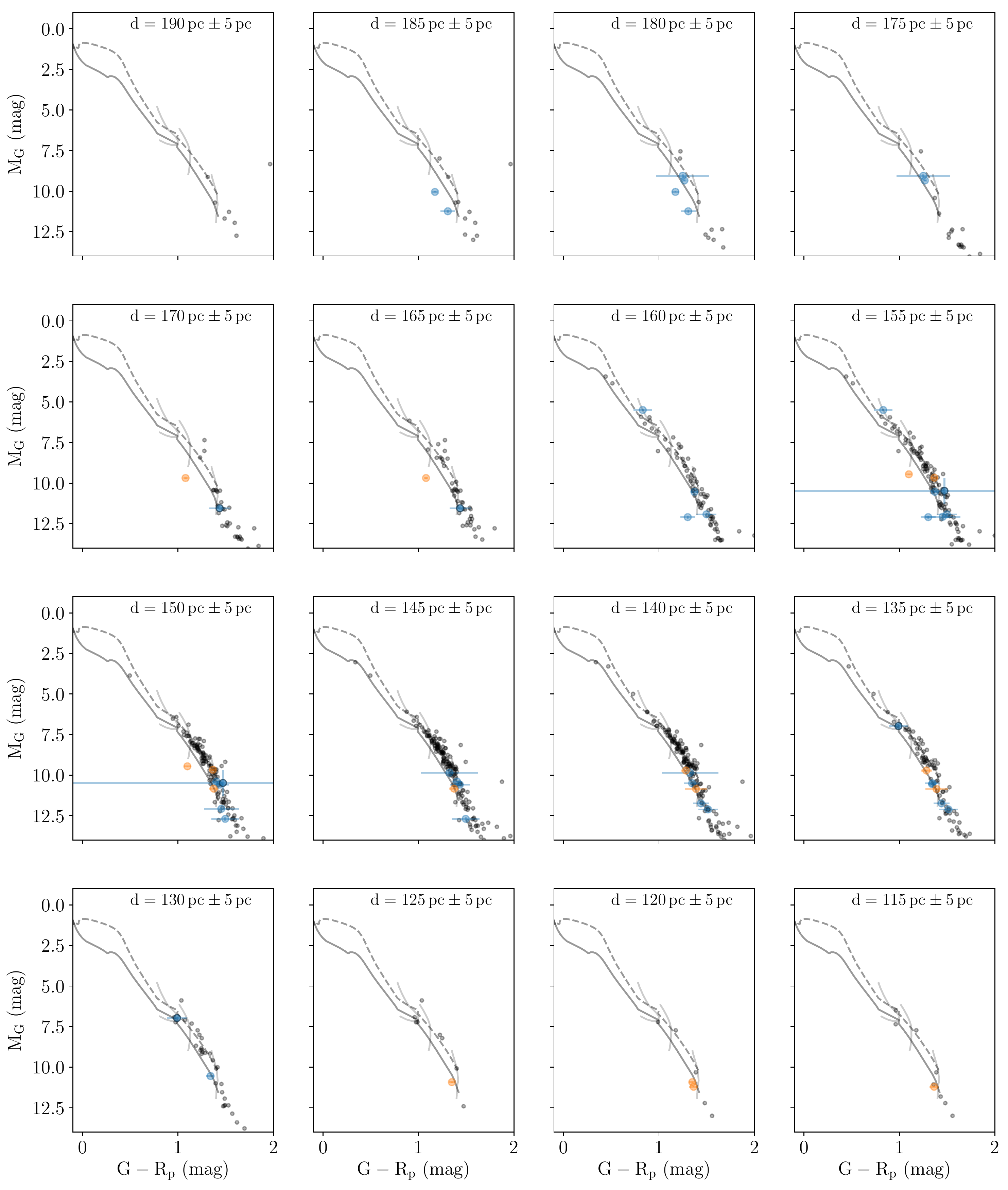}
\caption{Absolute magnitude vs. color for the previously known Upper Sco members (gray circles) and the new candidate members presented in this paper. The YSO symbols are the same as those shown in Figure \ref{fig:pm}. The extinction-free isochrones correspond to 4\,Myr (dashed line, dark gray) and 12\,Myr (solid line, dark gray) \citep{marigo17}, with mass tracks of 0.1\msun, 0.4\msun, and 0.7\msun\ for reference.}
\label{fig:cmd_US_gaia}
\end{figure*}

\clearpage

\renewcommand{\thefootnote}{\fnsymbol{footnote}}
\begin{longtable}{cccc}
\caption{\label{tab:new_US_ysos} List of new disk-bearing YSO US candidate members.}\\
\hline\hline
Gaia designation & ALLWISE designation  &
Disk type & notes\protect\footnotemark[1] \\
\hline
\endfirsthead
\caption{continued.}\\
\hline
Gaia designation & ALLWISE designation  &
Disk type & notes\protect\footnotemark[1] \\
\hline
\endhead
\hline
\endfoot
\hline
\multicolumn{4}{c}{New candidate members with disks}\\
\hline
Gaia DR2 6038919529389764352 & WISEA J162137.18-285833.0$^\dagger$ & Thick &  LR \\ 
Gaia DR2 6235477127144134144 & WISEA J155710.44-254531.0$^\dagger$ & Thick &  \\ 
Gaia DR2 6040977192383027328 & WISEA J154857.66-290900.3$^\dagger$ & Thick &  \\ 
Gaia DR2 6233762782358231808 & WISEA J155208.83-272345.9$^\dagger$ & Thick &  \\ 
Gaia DR2 6233788105485713408 & WISEA J155125.62-270743.4 & Thick &  \\ 
Gaia DR2 6036840215470943872 & WISEA J160710.28-312547.2$^\dagger$ & Thick &  \\ 
Gaia DR2 6236164837307632256 & WISEA J155832.63-243127.4$^\dagger$ & Thick &  LR \\ 
Gaia DR2 6236682741640391808 & WISEA J154854.33-244310.3$^\dagger$ & Thick &  \\ 
Gaia DR2 6236696073218711424 & WISEA J154853.82-244055.3 & Thick &  LR \\ 
Gaia DR2 6239076683392683008 & WISEA J152921.88-240305.1 & Thick &  \\ 
Gaia DR2 6240374008334811008 & WISEA J154925.69-225036.3 & Thick &  \\ 
Gaia DR2 6040286359786429568 & WISEA J155942.30-294549.7 & Thick &  Ovar \\ 
Gaia DR2 6037557131412384000 & WISEA J161932.29-303831.7$^\dagger$ & Anemic &  \\ 
Gaia DR2 6227034355309750016 & WISEA J151822.15-242328.1$^\dagger$ & Anemic &  \\ 
Gaia DR2 6235703076781985152 & WISEA J155855.10-251522.9 & Anemic &  \\ 
Gaia DR2 6036792283641120128 & WISEA J161029.83-304653.8$^\dagger$ & Anemic &  \\ 
Gaia DR2 6036986759757743232 & WISEA J160925.12-303409.5 & Anemic &  \\ 
Gaia DR2 6234946681499736320 & WISEA J154820.73-252503.4 & Anemic &  LR \\ 
Gaia DR2 6042477514657273728 & WISEA J161046.27-272144.2$^\dagger$ & Anemic &  \\ 

\hline
\multicolumn{4}{c}{Previously known members with disks}\\
\hline
Gaia DR2 6035606597786562048 & WISEA J161520.22-325505.3 & Thick &  \\ 
Gaia DR2 6038520956430662144 & WISEA J161206.67-301027.3 & Thick &  MIRvar \\ 
Gaia DR2 6235335049625674496 & WISEA J155432.47-262933.5$^\dagger$ & Thick &  \\ 
Gaia DR2 6235355661166851200 & WISEA J155703.49-261008.3$^\dagger$ & Thick &  MIRvar \\ 
Gaia DR2 6235774029641941888 & WISEA J155704.89-245522.9 & Thick &  \\ 
Gaia DR2 6234842567193557888 & WISEA J154725.71-260918.8 & Thick &  \\ 
Gaia DR2 6237048805996330624 & WISEA J155601.02-233808.3$^\dagger$ & Thick &  \\ 
Gaia DR2 6042963567523268480 & WISEA J160528.73-265550.0 & Thick &  \\ 
Gaia DR2 6043375025387317504 & WISEA J160728.62-263013.3$^\dagger$ & Thick &  \\ 
Gaia DR2 6043066337495355520 & WISEA J155823.75-272143.9 & Anemic &  \\ 

\footnotetext[0]{$\dagger$: possible source confusion}
\footnotetext[1]{Ovar: flagged as variable by Gaia DR2; MIRvar: flagged as variable by ALLWISE; LR: Gaia DR\,2 large RUWVE (defined as greater than 1.4)}
\end{longtable}

\clearpage

\section{V1062 Scorpii}
\label{sec:v1062sco}

\begin{figure*}[!h]
\centering
\includegraphics[width=\textwidth]{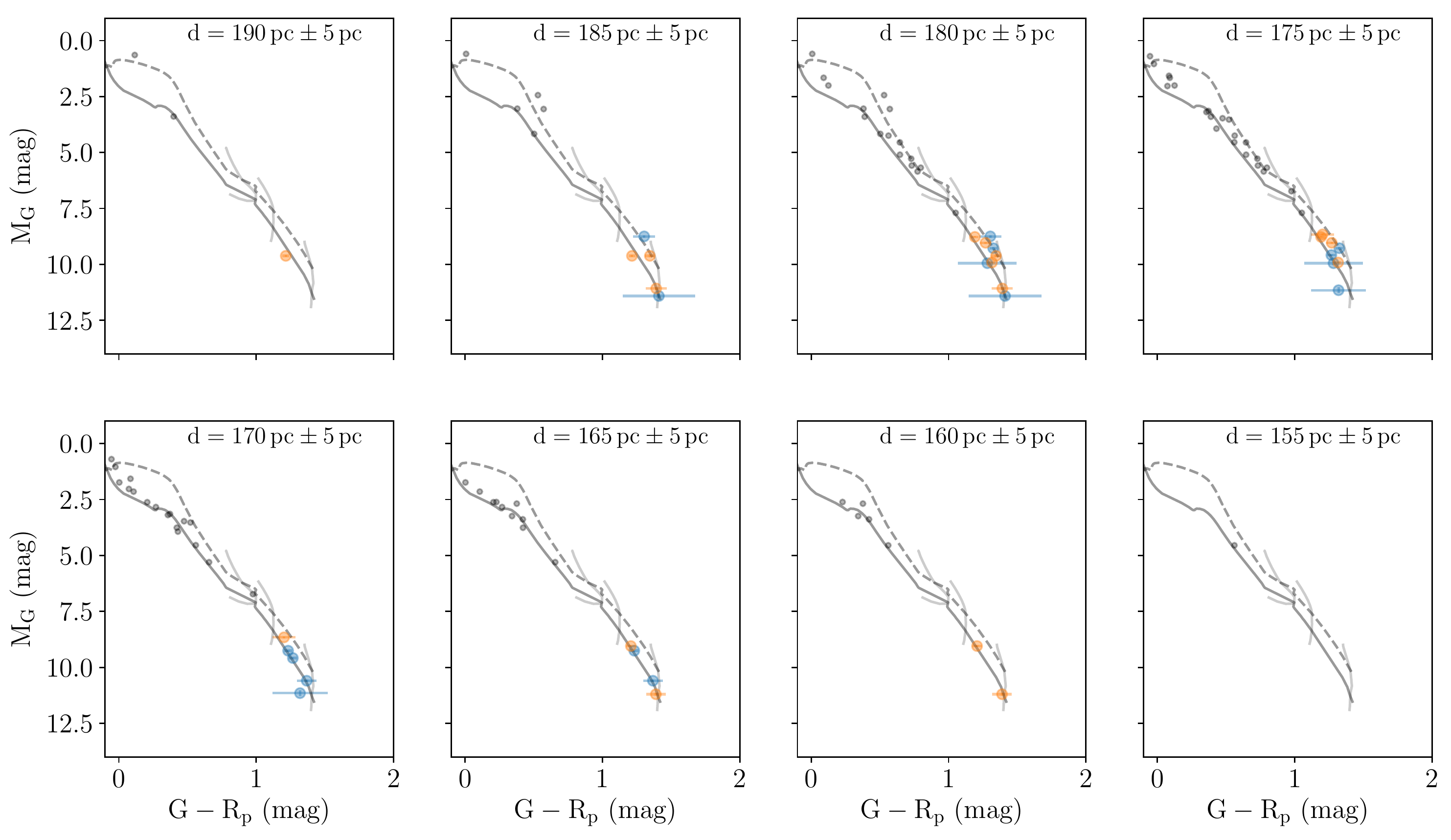}
\caption{Absolute magnitude vs. color for the previously known V1062 Sco members (gray circles) and the new candidate members presented in this paper. The YSO symbols are the same as those shown in Figure \ref{fig:pm}. The extinction-free isochrones correspond to 4\,Myr (dashed line, dark gray) and 12\,Myr (solid line, dark gray) \citep{marigo17}, with mass tracks of 0.1\msun, 0.4\msun, and 0.7\msun\ for reference.}
\label{fig:cmd_v1062sco_gaia}
\end{figure*}

\renewcommand{\thefootnote}{\fnsymbol{footnote}}
\begin{longtable}{cccc}
\caption{\label{tab:new_V1062sco_ysos} List of candidate new V1062 Sco disk-bearing YSOs}.\\
\hline\hline
Gaia designation & ALLWISE designation & 
Disk type & notes\protect\footnotemark[2] \\
\hline
\endfirsthead
\caption{continued.}\\
\hline
Gaia designation & ALLWISE designation & 
Disk type & notes\protect\footnotemark[2] \\
\hline
\endhead
\hline
\endfoot
\hline

Gaia DR2 5970057695957697408 & WISEA J165223.98-390553.2$^\dagger$ & Thick &  \\ 
Gaia DR2 6017441390077163904 & WISEA J163648.35-391133.6$^\dagger$ & Thick &  \\ 
Gaia DR2 5970467543251393792 & WISEA J165343.69-383003.2$^\dagger$ & Thick &  \\ 
Gaia DR2 6019330591570879872 & WISEA J164248.21-364800.6$^\dagger$ & Thick &  \\ 
Gaia DR2 6016952661465264128 & WISEA J163317.82-405107.8$^\dagger$ & Thick &  \\ 
Gaia DR2 6017434866048226944 & WISEA J163656.25-392223.2$^\dagger$ & Thick &  \\ 
Gaia DR2 6019195248562656128 & WISEA J164137.90-372351.3$^\dagger$ & Thick &  \\ 
Gaia DR2 5970467130934523264 & WISEA J165336.13-383138.3$^\dagger$ & Thick &  \\ 
Gaia DR2 5971240255071168128 & WISEA J165240.89-380550.6$^\dagger$ & Thick &  LR \\ 
Gaia DR2 6017488432884483072 & WISEA J163616.81-390608.0$^\dagger$ & Thick &  \\ 
Gaia DR2 5970833199566225152 & WISEA J165146.92-384645.3$^\dagger$ & Anemic &  \\ 
Gaia DR2 5969470728548177408 & WISEA J163930.22-393137.1 & Anemic &  \\ 
Gaia DR2 6017515405277725952 & WISEA J163846.97-391711.6$^\dagger$ & Anemic &  \\ 
Gaia DR2 5968662454388595840 & WISEA J163815.85-412340.2$^\dagger$ & Anemic &  \\ 
Gaia DR2 6017453798266557952 & WISEA J163510.95-392844.9$^\dagger$ & Anemic &  \\ 
Gaia DR2 6016970116212453248 & WISEA J163224.91-403803.6$^\dagger$ & Anemic &  \\ 
Gaia DR2 5971493284521616512 & WISEA J164940.10-374831.5$^\dagger$ & Anemic &  \\ 
Gaia DR2 6019205492082159232 & WISEA J164423.53-374321.9$^\dagger$ & Anemic &  \\ 
Gaia DR2 5968880604372020736 & WISEA J163648.50-405452.2$^\dagger$ & Anemic &  \\ 

\footnotetext[0]{$\dagger$: possible source confusion}
\footnotetext[1]{Ovar: flagged as variable by Gaia DR2; MIRvar: flagged as variable by ALLWISE; LR: Gaia DR\,2 large RUWVE (defined as greater than 1.4)}
\end{longtable}

\section{Non-parametric hypothesis testing of distance PDFs}

\setulcolor{cinza}
\setul{0.1ex}{0.5ex}

\begin{table}[!h]
    \caption{Summary of p-values for the different paired distance probability density functions (PDFs). The null hypothesis (i.e., the distributions are drawn randomly from the same population) is rejected for p-values less than 0.05 and are marked in gray (borderline values are underlined in gray).}
    \label{tab:p-values}
    \centering
    \begin{tabular}{c|ccc}
    \hline
    \hline
    & \multicolumn{3}{c}{p-values} \\
     Panel Figure \ref{fig:age_dist}  & Anderson-Darling & Kolmogorov–Smirnov & Wilcoxon-Mann-Whitney U  \\
    \hline
    (a) & 0.8894 & 0.9669 & 0.4979 \\
    (b) & \cellcolor{cinza} 0.0051 & \cellcolor{cinza}0.0002 & \cellcolor{cinza} 0.0025 \\
    (c) &  \ul{0.0503} & \cellcolor{cinza} 0.0271 & \cellcolor{cinza} 0.0410 \\
    \hline
    (d) & \cellcolor{cinza} 0.0187 & \ul{0.0593} & 0.3374 \\
    (e) & 0.1200 & 0.1798 & 0.0667 \\
    (f) & \cellcolor{cinza} 0.0004 & \cellcolor{cinza} 0.0025 & \cellcolor{cinza} 0.0002 \\
    \hline
    (g) & 0.6164 & 0.6516 & 0.4335 \\
    (h) & 0.4579 & 0.6885 & 0.1881\\
    (i) & 0.3814 & 0.4308 & 0.1522\\
    \hline
    (k) & 0.4320 & 0.4605 & 0.1328\\\
    (l) & 0.3626 & 0.4232 & 0.1533\\
    \hline
    \end{tabular}
\end{table}{}

\noindent The Anderson-Darling test is more sensitive to the tails of the distributions being compared, while the K-S test is more sensitive to the centers of the distributions. One should also use caution in the interpretation of the results due to the small number of sources in some distributions ($<$20).

\end{appendix}

\end{document}